\documentclass{svjour3}                     

\usepackage{graphicx}
\usepackage{amssymb,amsmath,bm}
\usepackage{color}

\usepackage{soul}

\newcommand {\Wi}{W\!i}

\journalname{Journal of Statistical Physics}

\begin{document}

\title{Subcritical instabilities in plane Poiseuille flow of an Oldroyd-B fluid
\thanks{A.M. acknowledges support from the UK Engineering and Physical Sciences Research Council (grant number EP/I004262/1).}
}

\author{Alexander Morozov         \and
        Wim van Saarloos
}


\institute{Alexander Morozov \at
              SUPA, School of Physics and Astronomy, The University of Edinburgh, James Clerk Maxwell Building, Peter Guthrie Tait Road, Edinburgh, EH9 3FD, United Kingdom \\
              Tel.: +44 131 650 5882\\
              \email{Alexander.Morozov@ed.ac.uk}           
           \and
           Wim van Saarloos \at
              Instituut--Lorentz, Universiteit Leiden, Postbus 9506, 2300 RA Leiden, The Netherlands \\
              \email{saarloos@lorentz.leidenuniv.nl} 
}

\date{Received: date / Accepted: date}

\maketitle

\begin{abstract}
Recently, detailed experiments on visco-elastic channel flow have provided convincing evidence for a nonlinear instability scenario which we had argued for based on calculations for visco-elastic Couette flow. Motivated by these experiments we extend the previous calculations to the case of visco-elastic Poiseuille flow, using the Oldroyd-B constitutive model. Our results confirm that the subcritical instability scenario is similar for both types of flow, and that the nonlinear transition occurs for Weissenberg numbers somewhat larger than one. We provide detailed results for the convergence of our expansion and for the spatial structure of the mode that drives the instability. This also gives insight into possible similarities with the mechanism of the transition to turbulence in Newtonian pipe flow.

\end{abstract}

\section{Introduction}
\label{intro}

Pierre Hohenberg played an important role in the scientific life and career of one of us, Wim van Saarloos, who was a junior colleague of Pierre at Bell Labs from 1982-1990. During these years Pierre worked steadily on his magnum opus on Pattern Formation, the review {\em Pattern formation outside of equilibrium} with Mike Cross, published in 1993 \cite{cross1993}. Although the two actually only started to work together during the last two of these years, indirectly and through informal discussions and by Pierre acting as a soundboard, Wim profited a lot from Pierre's wisdom and insight in pattern formation. Moreover, Pierre's attention for understanding real experiments, for translating one's theoretical analysis into experimentally testable predictions, and his emphasis on the importance of writing longer papers and reviews, has had a lasting influence on him. It is therefore a privilege to contribute to this special issue honoring Pierre Hohenberg, with a topic that has many elements of Pierre's interests and ways of doing physics.

In his personal reflections which will appear in a memorial book for Pierre Hohenberg \cite{vansaarloos2018}, Wim has extensively described Pierre's influence on him. Moreover, it is explained there in detail how the topic which the two of us, authors of this paper, explored together some ten to fifteen years ago, exhibits traces of  the earlier collaborations of Wim and Pierre. We refer to these personal reminiscences \cite{vansaarloos2018} for this background. Connections relevant to the present paper are: the relation with their unpublished explorations of the transition to turbulence in Newtonian pipe flow, how this culminated in their long paper on the quintic Complex Ginzburg-Landau equation \cite{vansaarloos1992},  how this in the end relates to the topic of interest here, and how much Pierre enjoyed it when we presented this whole storyline at the Rutgers meeting celebrating his 80$^{th}$ birthday in December 2014.

The issue at stake is the question whether viscoelastic channel flow exhibits a nonlinear flow-instability in the small Reynolds number limit.

It is well known that Newtonian fluids flowing through a pipe, so-called Poiseuil\-le flow, exhibit a nonlinear instability to turbulence \cite{Eckhardt2008,barkley2016}. The transition {\em must be nonlinear}, since the laminar Poiseuille flow state is linearly stable. The same holds for Couette flow, flow induced by having two plates moving in opposite direction.\footnote{Planar Poiseuille flow actually  does become linearly unstable at sufficiently high Reynolds numbers, but this linear instability is actually irrelevant for the transition to turbulence, which happens at much lower Reynolds numbers.} 

Adding polymers to a fluid can have drastic effects. With  polymers in it, the fluid behaves as a so-called viscoelastic fluid. Because shear causes stretching of the polymers, a viscoelastic polymer fluid exhibits elasticity, relaxation and anisotropy:  each stretched polymer acts like a little elastic rubber band which is oriented by the shear direction and which takes time to respond (relax) when the local shear rate varies. These effects are stronger, of course, the longer the polymer molecules are, and the higher their concentration. 
A  well-know dramatic manifestation of practical interest of these viscoelastic effects is 'turbulent drag reduction' \cite{White2008}:  when sufficiently long polymers are added in small amounts to a fluid, the drag experienced by turbulent flow through the pipe is reduced. The precise origin of this effect is still a matter of active research. 

We are here interested in another limit, the limit in which both the polymer concentration and their length are large enough that the effective fluid viscosity is large, so much so that the flow can be considered as small Reynolds number flow. The convective nonlinearity in the Navier-Stokes equation can then be neglected, and the only nonlinearities come from the so-called constitutive equation that relates the polymer stress tensor and the flow field. In this limit, the only dimensionless quantity for simple shear flows is the so-called Weissenberg number $\Wi$, which is a measure of the normal stress effects in the fluid, resulting from the orientation and stretching of the polymers.

It has been known since 1977 \cite{Ho1977} that viscoelastic Poiseuille flow, as modeled by the so-called Oldroyd-B constitutive equation, is linearly stable in the small Reynolds number limit. Motivated by experimental work by  Bonn and co-workers \cite{Bertola2003,Bonn2011}, and by the similarity of the questions concerning the nonlinear transition to turbulence of Newtonian fluids, we therefore explored theoretically in a series of papers \cite{Meulenbroek2003,Meulenbroek2004,Morozov2005prl,Morozov2007} the question whether parallel viscoelastic shear flows would show a nonlinear flow instability as well. 

A combination of analytical arguments, based on insights into  the Newtonian case and the theory of dynamical systems and pattern formation \cite{Morozov2007}, and explicit but approximate nonlinear stability calculations for the case of Couette flow, led us to propose \cite{Bertola2003,Morozov2005prl,Morozov2007} that  viscoelastic shear flows should indeed show this putative nonlinear transition for Weissenberg numbers somewhat larger than one. We speculated that the transition would lead to turbulent flow.

Several initial attempts to test our proposed scenario experimentally gave negative results. Some numerical investigations did show behavior reminiscent of the proposed scenario, but it is well known to  experts that simulations of these bilinear type of constitutive equations are very prone to numerical instabilities, resulting from flow regions which lead to exponential divergence of stress components.Thus, also numerical investigations were considered to give inconclusive results.  

The issue therefore remained unsettled till Arratia and co-workers \cite{Pan2013,Qin2017} convincingly showed in 2013, using microfluidics experiments, that indeed small Reynolds number viscoelastic channel flow  {\em does} indeed exhibit a nonlinear transition to turbulence. Qualitatively, their finding is very much in line with what we had proposed: by perturbing the viscoelastic flow in very long micro-channels behind the inlet in a  controlled way, by putting a variable number of cylindrical obstacles in the channel, they found that sufficiently strong perturbations lead to a well-developed asymptotic turbulent state far down the channel. Moreover, the experiments showed that the transition occurs for Weissenberg numbers somewhat larger than one, very much like what our own approximate analysis had suggested. 

The unequivocal experimental evidence for the nonlinear viscoelastic instability of channel flow leads us to revisit and extend, in this paper, our earlier theoretical analysis. The earlier calculations \cite{Morozov2005prl} were focussed on Couette flow (two plates moving in opposite directions); we here extend the results to the experimentally relevant case of Poiseuille flow. We also take the opportunity to present in more detail the nature of our analysis, which is based on an asymptotic Amplitude-equation-like expansion taken to unusually large order (which, incidentally, was an aspect that Pierre liked very much when we presented this at the Rutgers meeting celebrating his 80th birthday). This also allows us to put our results into perspective and to substantiate our previous claim that our results converge to well-defined values and flow profiles. Indeed, we also show for the first time the spatial structure of the nonlinear modulated waves which according to our analysis determine the instability threshold.

In the next section, we simply summarise the main result of our analysis, which shows within the context of the same approximate analysis  that viscoelastic Poiseuille flow between plates exhibits a nonlinear instability which is very similar to the one for plane Couette flow published previously \cite{Morozov2005prl}. Also the transition amplitude and the values of the Weissenberg number where the instability is found, is similar. In section \ref{sect:AmplEq} we then present the technical details of the expansion underlying our results, which we then apply to the case of plane Poiseuille flow in section \ref{sect:results}. In our concluding section \ref{sect:discussion} we put our results into perspective, and speculate on the transition to turbulence mechanism of viscoelastic flow.

\begin{figure}[htp]
\centering
\includegraphics[width=0.65\textwidth]{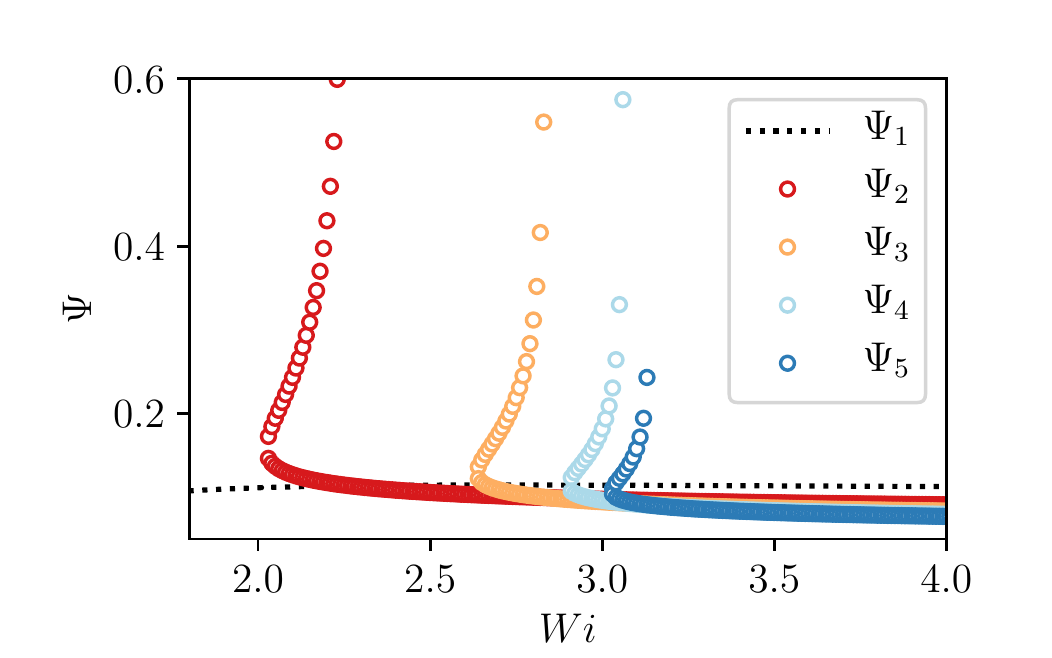}
\caption{Steady-state amplitude of the travelling-wave solution in plane Couette flow of UCM fluids found in \cite{Morozov2005} for $k_x=k_z=1$; the ratio $Re/\Wi=10^{-3}$ was kept constant. Re-plotted from Ref.\cite{Morozov2005}. For definitions see Section \ref{sect:UCM}.}
\label{Fig:PCbifurcation}
\end{figure}

\section{Main result of perturbative expansion for viscoelastic plane Couette and plane Poiseuille flow}
\label{sect:summary}

Figure 1 summarizes our earlier result \cite{Morozov2005prl} for plane Couette flow and our extensions to the case of Poiseuille flow between plates (right panel) as a function of increasing Weissenberg number $\Wi$ which, as stated above, is the only dimensionless number for these flow configurations in the zero Reynolds number limit. What do these results mean?

As detailed in later sections, these calculations are based on perturbing the laminar viscoelastic flow with a finite-wavelength mode of amplitude $\Phi$. For flow between plates, one does this by picking a wavenumber $k_x$ associated with the wavenumber of the modulation along the flow direction, and a wavenumber $k_z$ in the transverse direction. We imagine these to be fixed, and then ask ourselves whether, if we think of the flow to be perturbed by a modulation of this type and of a given initial amplitude $\Phi$, this amplitude would decay --- meaning stability --- or increase in time (meaning instability). The linear stability calculation, already done long ago \cite{Ho1977}, is based on assuming that the mode amplitude $\Phi$ (for a given $k_x$ and $k_z$) is infinitesimally small, and the finding from these calculations is that when the initial amplitude $\Phi$ is infinitesimally small, it will indeed decay in time: the flow is {\em linearly} stable.

Our calculations are based on probing whether $\Phi$ will grow or decay in time perturbatively, by performing an expansion in powers of $\Phi$, for fixed wavenumbers. In other words, we only probe growth or decay of the amplitude, without allowing for spatial instabilities of the modes themselves. 

In Figure 1, we plot along the vertical axis the critical dimensionless value $\Psi=|\Phi_c|$ of the amplitude --- which we can think of the the maximum relative change in shear rate  at the walls --- separating decay (for smaller amplitudes) from growth (for larger amplitudes) for particular values of the wavenumbers indicated. 

Clearly, these results indicate that within the limitations of our perturbative calculations, we expect that nonlinear instability to be very similar for plane Couette and Poiseuille flow. This is of course what one would expect physically as the instability is driven by the self-amplifying nature of viscoelastic flow along curved streamlines. This should depend little on the details of the way the flow is driven.

Of course, as explained above our analysis is based on an amplitude expansion, which itself is an asymptotic expansion. It is legitimate to ask why we were confident  to draw conclusions from such an expansion. To illustrate why we did dare to do so, we show in Figure \ref{Fig:PCbifurcation}  the results up to 5$^{th}$ (red, label $\Psi_2$), 7$^{th}$ (orange, label $\Psi_3$) , 9$^{th}$ (light blue, label $\Psi_4$) and even 11$^{th}$ order (dark blue, label $\Psi_5$)  in our expansion for $\Phi_c$ for the case of plane Couette flow. Quite surprisingly --- we had no reason to expect this a priori --- the results for the critical value $\Phi_c$ are quite robust. Moreover, even the 'nose' of the curves on the left seems to converge: whenever the expansion is extended by including another order, the curve shifts by about half the amount that it did in the previous step.

\section{Derivation of the amplitude equation}
\label{sect:AmplEq}
In this Section we present a method of approximating non-linear solutions for a class of non-linear partial differential equations in the following form
\begin{equation}
\label{MainEq}
\hat{\mathcal L}V + \hat{A}\frac{\partial V}{\partial t} = N\left(V,V\right),
\end{equation}
where $V$ is a $d$-dimensional vector of fields, $\hat{\mathcal L}$ and  $\hat{A}$ are linear operators, and $N$ is a quadratic non-linear operator. Our method is primarily designed for problems of hydrodynamic stability in parallel shear flows, and, from the onset, we incorporate some of the main features of the corresponding equations of motion directly into Eq.\eqref{MainEq}. First, we introduce a Cartesian coordinate system $(x,y,z)$, where $x$ and $z$ denote the unbounded directions, while $y$ is the direction between two confining plates. The vector $V$ is thought of to contain the velocity, stress components, and pressure in the fluid. We further assume that $\hat{\mathcal L}$, $\hat{A}$, and $N$ only contain spatial derivatives, and that $N\left(f(t)V_1,V_2\right) = N\left(V_1,f(t)V_2\right)=f(t)N\left(V_1,V_2\right)$, where $V_1$ and $V_2$ are arbitrary functions of space, while $f(t)$ is an arbitrary function of time.

Our goal is to demonstrate that Eq.\eqref{MainEq} can have finite-amplitude solutions besides the trivial (laminar) one. In the absence of a linear instability, there is no systematic way of finding such a solution and we attempt to construct it perturbatively, in a way motivated by amplitude expansion methods developed for studying near-threshold behaviour of pattern-forming instabilities (see \cite{cross1993,vanHecke1994}, for example). Our expansion is based on the eigenfunctions, $e^{i\left(k_x x + k_z z\right)}V_0^{(n)}(y)$, of the linear operator:
\begin{equation}
\label{linear_eigenmode}
\hat{\mathcal L} e^{i\left(k_x x + k_z z \right)}V_0^{(n)}(y)=-\lambda_n \hat{A}e^{i\left(k_x x + k_z z \right)}  V_0^{(n)}(y).
\end{equation}
The index $n$ enumerates the eigenfunctions; their number and form depends on the problem at hand. 
The desired finite-amplitude solution for the real-valued fields is then \emph{assumed} to be represented by
\begin{align}
\label{solution_form}
& V(x,y,z,t) = \Phi(t)e^{i\xi}V_0(y) + \Phi^{*}(t)e^{-i\xi}V_0^{*}(y) \nonumber \\
& \, + U_0(y,t) + \sum_{n=2}^{\infty}\left[ U_n(y,t)e^{i n\xi} + U_n^{*}(y,t)e^{-i n\xi}\right],
\end{align}
where $e^{i\xi}V_0$ is one of the eigenfunctions, and we have introduced $\xi=k_x x + k_z z$; ''$*$'' denotes complex conjugation; the choice of a particular eigenfunction will be discussed below, but we note here that for plane Couette flow these are the Gorodtsov-Leonov modes \cite{Gorodtsov1967} that we used in our previous work \cite{Morozov2005prl}.  One can view Eq.\eqref{solution_form} as a Fourier expansion in $x$ and $z$ with an extra assumption about the form and interrelation between the coefficients; $\Phi(t)$ is the basic amplitude of the mode whose non-linear behaviour we aim to study perturbatively. In the spirit of amplitude-equation techniques \cite{cross1993,vanHecke1994}, the time-dependent amplitude $\Phi(t)$ follows the linear dynamics governed by Eq.\eqref{linear_eigenmode} at short times, which is replaced by nonlinear dynamics at large $t$. In the following, we derive the evolution equation for the amplitude $\Phi(t)$ assuming that it is small in some sense. This assumption will be checked for self-consistency, once the solution is obtained.

To proceed, we assume that the dominant dynamics in Eq.\eqref{solution_form} are given by the time-evolution of $\Phi(t)$, and that the higher harmonics $U_n$'s follow it adiabatically (they are 'slaved' to $\Phi(t)$). This implies that the Fourier components $U_n$'s can only arise as a result of nonlinear interactions of the eigenmode with itself and other $U_n$'s. For instance, if we denote $V_l = \Phi(t)e^{i\xi}V_0(y)$, then $U_0(y)$ has contributions from the interactions of a) $V_l$ with $V_l^*$, 
b) $V_l$, $V_l^*$, $V_l$, and $V_l^*$, etc. Applying this power-counting argument to all Fourier modes, we obtain
\begin{align}
&U_0(y,t)=|\Phi(t)|^2 u_0^{(2)}(y) + |\Phi(t)|^4 u_0^{(4)}(y) + \cdots , \nonumber \\
&U_2(y,t)= \Phi^2(t) u_2^{(2)}(y) + \Phi^2(t)|\Phi(t)|^2 u_2^{(4)}(y) + \cdots ,\label{Un} \\
&U_3(y,t)= \Phi^3(t) u_3^{(3)}(y) + \cdots , \nonumber \\
&\cdots, \nonumber
\end{align} 
where $u_n^{(m)}(y)$ are unknown functions, which will be determined below; the subscript $n$ and superscript $m$ correspond to the Fourier harmonic and the order in $\Phi$, respectively.

To derive the time-evolution equation for $\Phi(t)$, we substitute Eq.\eqref{solution_form} into Eq.\eqref{MainEq} and separate the terms proportional to $e^{i\xi}$. This yields
\begin{align}
&\hat{\mathcal L}\left(\Phi(t)e^{i\xi}V_0(y)\right) + \frac{d\Phi(t)}{dt} \hat{A} \left(e^{i\xi}V_0(y)\right)= 
\left( \frac{d\Phi(t)}{dt} - \lambda \Phi(t)\right) \hat{A} \left(e^{i\xi}V_0(y)\right)  \nonumber \\
& = \bar{N}\left(\Phi(t)e^{i\xi}V_0(y), U_0(y,t)\right) + \bar{N}\left( \Phi^{*}(t)e^{-i\xi}V_0^{*}(y), U_2(y,t)e^{2 i \xi}\right) \nonumber\\
& \qquad\qquad + \sum_{n=2}^{\infty} \bar{N} \left( U_{n+1}(y,t)e^{i (n+1)\xi},U_n^{*}(y,t)e^{-i n\xi}\right),
\label{substituted} 
\end{align}
where we used Eq.\eqref{linear_eigenmode} and introduced $\bar{N}\left(A,B\right) = N\left(A,B\right) + N\left(B,A\right)$. Although the eigenmodes of the linear operators involved in flow stability problems often form full sets, they are typically non-normal \cite{SchmidHenningson2000} and their eigenmodes are not orthogonal. Therefore, to calculate the contribution of the r.h.s. of Eq.\eqref{substituted} to $d\Phi/dt$, we employ the adjoint operator $\hat{\mathcal{L}}^\dagger$, defined via
\begin{equation}
\label{definition_adjoint}
\langle V_1 \vert \hat{\mathcal{L}} V_2 \rangle = \langle \hat{\mathcal{L}}^\dagger V_1 \vert V_2 \rangle ,
\end{equation}
where $V_1$ and $V_2$ are two \emph{arbitrary} vectors satisfying proper boundary conditions.  The scalar product $\langle \cdots\rangle$ is given by
\begin{align}
\langle V_1 \vert V_2 \rangle = \lim_{L_x,L_z\rightarrow \infty} \frac{1}{2 L_x} \int_{-L_x}^{L_x} dx \, \frac{1}{2d}\int_{-d}^{d} dy \, \frac{1}{2 L_z} 
\int_{-L_z}^{L_z} dz \, \left( V_1^*,V_2 \right),
\end{align}
where $\left(A,B\right)=\sum_i A_i B_i$ is the Euclidean dot product, and $2d$ is the distance between the confining plates. The actual form of the adjoint operator is obtained by performing integration-by-parts in Eq.\eqref{definition_adjoint}, as will be demonstrated later. The eigenmodes $e^{i\left(k_x x + k_z z\right)} W_0^{(m)}(y)$ and the eigenvalues $\chi_m$ of the adjoint operator are given by
\begin{equation}
\label{adjoint_eigenfunctions}
\hat{\mathcal L}^\dagger e^{i\left(k_x x + k_z z\right)} W_0^{(m)}(y) = -\chi_m \hat{A}^\dagger e^{i\left(k_x x + k_z z\right)} W_0^{(m)}(y),
\end{equation}
with $m=1,2,\dots$, and $\hat{A}^\dagger$ being the operator adjoint to $\hat{A}$; for the problem we consider in Section \ref{sect:UCM}, operator $\hat{A}$ is self-adjoint. Since Eq.\eqref{linear_eigenmode} is a generalised eigenvalue problem, the orthogonality condition between the eigenmodes of the linear and adjoint operators is somewhat unusual, and we state it here explicitly. We consider 
\begin{align}
 \langle e^{i\xi} W_0^{(m)}(y) \vert \hat{\mathcal L} e^{i\xi} V_0^{(n)}(y)\rangle = -\lambda_n \langle e^{i\xi} W_0^{(m)}(y) \vert \hat{A} e^{i\xi} V_0^{(n)}(y)\rangle.
\label{adp1}
\end{align}
At the same time, 
\begin{align}
&\qquad \qquad \langle e^{i\xi} W_0^{(m)}(y) \vert \hat{\mathcal L} e^{i\xi} V_0^{(n)}(y)\rangle = \langle \hat{\mathcal L}^\dagger e^{i\xi} W_0^{(m)}(y) \vert e^{i\xi} V_0^{(n)}(y)\rangle \nonumber \\
&= -\chi_m^* \langle\hat{A}^\dagger e^{i\xi} W_0^{(m)}(y) \vert e^{i\xi} V_0^{(n)}(y)\rangle 
= -\chi_m^* \langle e^{i\xi} W_0^{(m)}(y) \vert \hat{A} e^{i\xi} V_0^{(n)}(y)\rangle.
\label{adp2}
\end{align}
Comparing Eqs.\eqref{adp1} and \eqref{adp2} we conclude that $\langle e^{i\xi} W_0^{(m)}(y) \vert \hat{A} e^{i\xi} V_0^{(n)}(y)\rangle = 0$, unless $\lambda_n = \chi_m^*$.

The evolution equation for the amplitude $\Phi(t)$ is finally obtained by projecting Eq.\eqref{substituted} onto the eigenmode of the adjoint operator $e^{i\xi} W_0(y)$, selected such that its eigenvalue $\chi=\lambda^*$. According to Eq.\eqref{Un}, the r.h.s. of Eq.\eqref{substituted} is a polynomial in $\Phi(t)$ and its complex conjugate, and one obtains
\begin{align}
\label{amplitude_equation}
&\frac{d\Phi}{d t} = \lambda \Phi + C_3 \Phi |\Phi|^2 + C_5 \Phi |\Phi|^4 + C_7 \Phi |\Phi|^6 + C_9 \Phi |\Phi|^8 + C_{11} \Phi |\Phi|^{10} \cdots.
\end{align}
The coefficients $C$'s are calculated by collecting terms of the corresponding order in $\Phi$ on the r.h.s. of Eq.\eqref{substituted} and projecting them on $e^{i\xi} W_0(y)$. Once these coefficients are known, we can then study whether there is a critical value $\Phi_c$ that separates decaying amplitudes, for small $|\Phi|$, from the growing ones.

To illustrate the method, we show here how to calculate the coefficient $C_3$, while the expressions for the higher-order coefficients are deferred to Appendix \ref{appendix_C}. To $O\left(\Phi |\Phi|^2\right)$, we obtain from Eq.\eqref{substituted} by projecting it on $e^{i\xi} W_0(y)$
\begin{align}
& C_3 = \frac{1}{\Delta}\biggl\langle e^{i\xi} W_0(y) \biggl\vert \bar{N}\left(e^{i\xi} V_0(y),u_0^{(2)}(y)\right) + \bar{N}\left(e^{-i\xi} V_0^*(y),e^{2 i\xi} u_2^{(2)}(y)\right)\biggr\rangle,
\label{C3}
\end{align}
where
\begin{align}
\Delta = \langle e^{i\xi} W_0(y) \vert e^{i\xi} \hat{A} V_0(y)\rangle.
\end{align}
Let us now determine the unknown functions $u_0^{(2)}$ and $u_2^{(2)}$. Once again, we substitute Eq.\eqref{solution_form} into Eq.\eqref{MainEq},
and separate the terms proportional to $e^{2 i \xi}$. To lowest order in $\Phi$, it yields
\begin{align}
\hat{\mathcal L}\left( \vert\Phi(t)\vert^2 u_0^{(2)}(y)\right) + \frac{d \vert\Phi(t)\vert^2 }{dt} \hat{A} u_0^{(2)}(y) = 
 \bar{N}\left(\Phi(t)e^{i\xi}V_0(y), \Phi^{*}(t)e^{-i\xi}V_0^{*}(y) \right).
\end{align}
The time-derivative can be evaluated self-consistently with the help of the amplitude equation (\ref{amplitude_equation}), and to $O\left(\Phi^2\right)$ is equal to
\begin{align}
\frac{d \vert\Phi\vert^2 }{dt} = \frac{d\Phi}{dt} \Phi^* + \Phi \frac{d\Phi^*}{dt} = \left( \lambda + \lambda^* \right) \vert\Phi\vert^2.
\end{align}
Therefore, $u_0^{(2)}(y)$ satisfies the following inhomogeneous ODE
\begin{align}
\hat{\mathcal L} u_0^{(2)}(y) + \left(\lambda + \lambda^* \right)\hat{A} u_0^{(2)}(y) = \bar{N}\left(e^{i\xi}V_0(y), e^{-i\xi}V_0^{*}(y) \right).
\label{u02}
\end{align}
Similarly, $u_2^{(2)}(y)$ is given by
\begin{align}
\hat{\mathcal{L}}\left( e^{2 i \xi} u_2^{(2)}\right) + 2\lambda  \hat{A} e^{2 i \xi} u_2^{(2)} = N\left( e^{i \xi}V_0,e^{i \xi}V_0\right). 
\label{u22}
\end{align}
The equations for higher $C$'s and $u$'s are obtained by a straightforward, though lengthy, generalisation to higher orders in $\Phi$ of the procedure outlined above. The corresponding expressions can be found in Appendix \ref{appendix_C}.

Equation \eqref{amplitude_equation}, together with the procedure to systematically determine the coefficients $C$'s, is the central result underlying our non-linear analysis of the flow stability. It allows one to calculate the amplitude of a non-linear solution to Eq.\eqref{MainEq} in the form given by Eq.\eqref{solution_form}. In what follows, we will be particularly interested in simple solutions to this equation, either in the form of stationary points or travelling waves. As we will see below, for the problem discussed in Section \ref{sect:UCM}, the least stable eigenvalues $\lambda$ are complex, and, therefore, the relevant solutions are of the latter type, given by $\Phi(t) = \Psi\,e^{i\,\Omega\,t}$, where $\Psi$ and $\Omega$ are real numbers. 
Substituting this ansatz into the amplitude equation \eqref{amplitude_equation}, and separating the real and imaginary parts, we obtain for a non-trivial solution with a stationary amplitude $\Psi$ 
\begin{align}
\label{threshold}
&0=Re \left (\lambda\right) + Re\left(C_3\right) \Psi^2 + Re\left(C_5\right) \Psi^4 + \cdots,\\
&\Omega = Im \left (\lambda\right) + Im\left(C_3\right) \Psi^2 + Im\left(C_5\right) \Psi^4 + \cdots.
\label{wavespeed}
\end{align}
The asymptotic nature of Eqs.\eqref{threshold} and \eqref{wavespeed} imply that only converging series can represent a physical solution. In turn, this translates into the requirement that the solution amplitude $\Phi(t)$ is sufficiently small, where the scale is given by the coefficients $C_n$. To study convergence of series Eqs.\eqref{threshold} and \eqref{wavespeed}, we employ a somewhat intuitive method based on the partial sums $S_m$, defined through 
\begin{align}
S_m \equiv Re \left (\lambda\right) + Re\left(C_3\right) \Psi^2 + \cdots + Re \left(C_{2m+1}\right) \Psi^{2m}.
\label{Sms}
\end{align}
We then solve a series of algebraic equations $S_1=0$, $S_2=0$, $\dots$, and obtain a corresponding series of solutions $\Psi_1$, $\Psi_2$, $\dots$, and $\Omega_1$, $\Omega_2$, $\dots$, using Eq.\eqref{wavespeed}. If both series approach limiting values, the latter are recognised as representing a physical solution; see Section \ref{sect:summary} for discussion.

In the next Sections we adopt this method to parallel shear flows of model viscoelastic fluids and calculate the coefficients in Eq.\eqref{amplitude_equation} up to $C_{11}$.

\section{Channel flow of a viscoelastic fluid}
\label{sect:UCM}

Here we adopt the method presented in the previous Section to the study of non-linear stability of parallel shear flows of model polymer fluids. We consider a flow between two parallel plates forced by either a constant pressure gradient $-\Delta P$ (plane Poiseuille flow). As in Section \ref{sect:AmplEq}, we select a coordinate system with $(x,y,z)$ being along the streamwise, vertical (gradient) and spanwise directions, respectively; the distance between the plates is $2 d$.

The velocity of the fluid $\bm v$ is governed by the Stokes equation
\begin{align}
\label{Stokes}
-\nabla p + \eta_s \nabla^2{\bm v} + \nabla\cdot{\bm \tau}  = 0,
\end{align}
and the incompressibility condition
\begin{align}
\label{incomp}
\nabla\cdot {\bm v} = 0,
\end{align}
where $p$ is the pressure, $\bm \tau$ is the polymeric contribution to the total stress, and $\eta_s$ is the solvent viscosity. In Eq.\eqref{Stokes}, we neglected the inertial terms as typical experiments on  elastic instabilities and turbulence are usually performed either in microfluidic devices or with high-viscosity fluids (see \cite{Groisman2004,Pan2013}, for example). In both cases, the corresponding values of the Reynolds number (defined as $Re=\mathcal{U}\,d/\nu$, where $\mathcal{U}$ is the typical flow velocity and $\nu$ is the kinematic viscosity of the fluid) do not exceed $10^{-2} - 10^{-1}$, and the inertial effects can be ignored.  

To describe the dynamics of the polymer stress tensor $\bm \tau$, we employ the Oldroyd-B model given by
\begin{align}
\label{Oldroyd}
&{\bm\tau} + \lambda\left[ \frac{\partial {\bm\tau}}{\partial t} + {\bm v}\cdot\nabla{\bm\tau} - \left(\nabla {\bm v}\right)^T\cdot{\bm\tau} - {\bm\tau}\cdot\left(\nabla {\bm v}\right)\right] = \eta_p \left[ \left(\nabla {\bm v}\right) + \left(\nabla {\bm v}\right)^T \right],
\end{align}
where $\lambda_M$ is the Maxwell relaxation time of the fluid, $\eta_p$ is the polymer viscosity, and $T$ denotes the matrix transpose. The Oldroyd-B model is the simplest equation incorporating normal-stress effects that are the driving force for many viscoelastic flow instabilities \cite{Larson1992,Shaqfeh1996,Morozov2007}; for a detailed discussion of various viscoelastic equations of motion and their predictions see \cite{Bird1987,Larson1988book,Morozov2015}.
Finally, the velocity field is assumed to satisfy the no-slip boundary condition
\begin{align}
\label{boundary_conditions}
v_x(x,y=\pm d,z) = v_y(x,y=\pm d,z) = v_z(x,y=\pm d,z) =0.
\end{align}

Equations \eqref{Stokes}-\eqref{boundary_conditions} have as laminar solution the well-known parabolic profile with ${\bm v}=\left(\mathcal{U}_0(y),0,0\right)$, where
\begin{eqnarray}
\label{U0_PPF}
\mathcal{U}_0(y) = \frac{\Delta P d^2}{2\left(\eta_s+\eta_p\right)}\left[ 1-\left(\frac{y}{d}\right)^2\right].
\end{eqnarray}
The corresponding elastic stresses are given by
\begin{align}
& \tau^{(0)}_{xx} = 2 \eta_p \lambda_M \mathcal{U}_0'^2(y), \nonumber \\ 
& \tau^{(0)}_{xy} = \eta_p \mathcal{U}_0'(y), \\
& \tau^{(0)}_{xz} = \tau^{(0)}_{yy } = \tau^{(0)}_{yz} = \tau^{(0)}_{zz} = 0. \nonumber
\end{align}

In what follows we render equations \eqref{Stokes}-\eqref{boundary_conditions} dimensionless by re-scaling the variables by appropriately chosen units, suggested by the laminar solution above.
We use $d$ as a unit of length, $\mathcal{U}=\Delta P d^2/(2\left(\eta_s+\eta_p\right))$ as a unit velocity, $d/\mathcal{U}$ as a unit of time, and $\eta_p\,\mathcal{U}/d$ as a unit of stress.
The material
properties of the fluid are controlled by two dimensionless parameters: the ratio of the solvent to total viscosities $\beta = \eta_s/(\eta_s+\eta_p)$, and the Weissenberg number $\Wi=\lambda_M \mathcal{U} / d$. The latter controls the strength of non-Newtonian effects in Eq.\eqref{Oldroyd} and plays in elastic instabilities and turbulence the same role as the Reynolds number does in Newtonian fluid mechanics.

Next, we split the dimensionless velocity and stress fields into the laminar part and a deviation from the laminar solution,
\begin{align}
& \tau_{ij} = 2\,\Wi\,\mathcal{U}_0'^2(y)\delta_{ix}\delta_{jx} + \mathcal{U}_0'(y)\left(\delta_{ix}\delta_{jy} + \delta_{iy}\delta_{jx}\right) + \tau^{(1)}_{ij}, \label{stresspertub} \\
& \bm{v} = \left( \mathcal{U}_0(y),0,0\right) + \bm{v}^{(1)},
\label{velpertub}
\end{align}
and introduce a perturbation vector $V=\left(\bm{v}^{(1)},{\bm \tau}^{(1)},p^{(1)}\right)^T$, where $p^{(1)}$ is the pressure perturbation. Substituting these expressions into the equations of motion, we arrive at the compact form, Eq.\eqref{MainEq}, introduced in Section \ref{sect:AmplEq}.
The explicit expressions for $\hat{\mathcal L}$, $N$ and $\hat{A}$ are given in Appendix \ref{appendix_A}. In the next Section we present the finite-amplitude solutions of Eqs.\eqref{Stokes}-\eqref{boundary_conditions} found by the method presented in Section \ref{sect:AmplEq}.

\section{Results}
\label{sect:results}

As the eigenmodes of the linear operator in Eq.\eqref{linear_eigenmode} are the starting point of our analysis, here we briefly present the linear stability analysis of plane Poiseuille flow of an Oldroyd-B fluid; for a detailed discussion see Wilson \emph{et al.} \cite{Wilson1999}. To calculate the eigenspectrum, we discretise Eq.\eqref{linear_eigenmode} with $\hat{\mathcal L}$ and $\hat{A}$ given in Appendix \ref{appendix_A} 
using the fully spectral Chebyshev-tau method \cite{Canuto:book}, and solve numerically the resulting generalised eigenvalue problem using Scientific Python \cite{scipy}.

\begin{figure}[h]
\hspace{0.5cm}
\includegraphics[width=0.48\textwidth]{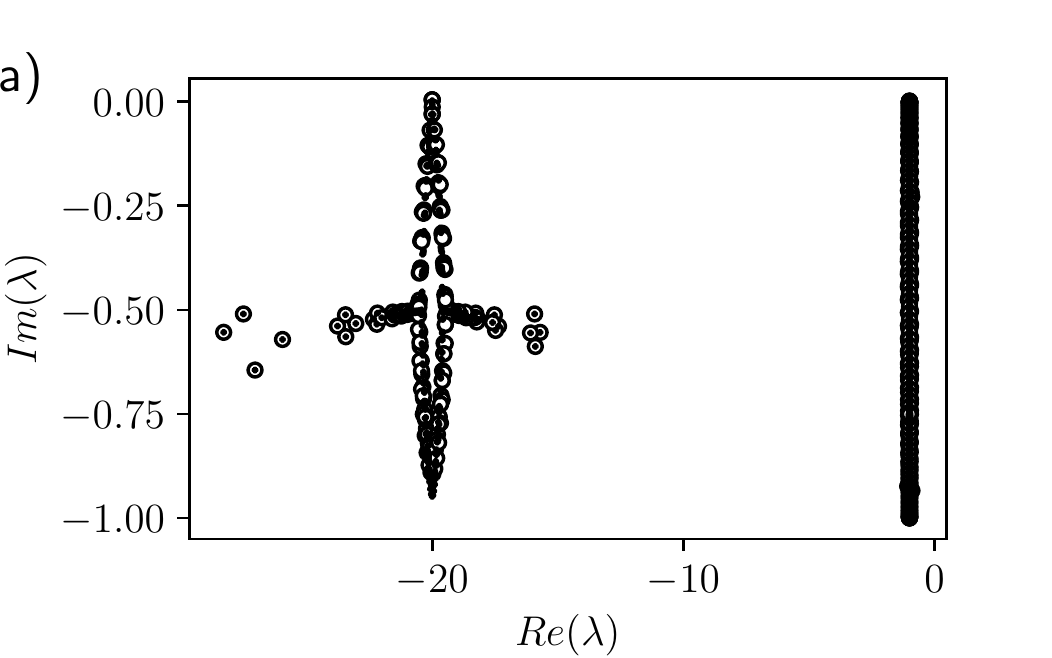}
\includegraphics[width=0.48\textwidth]{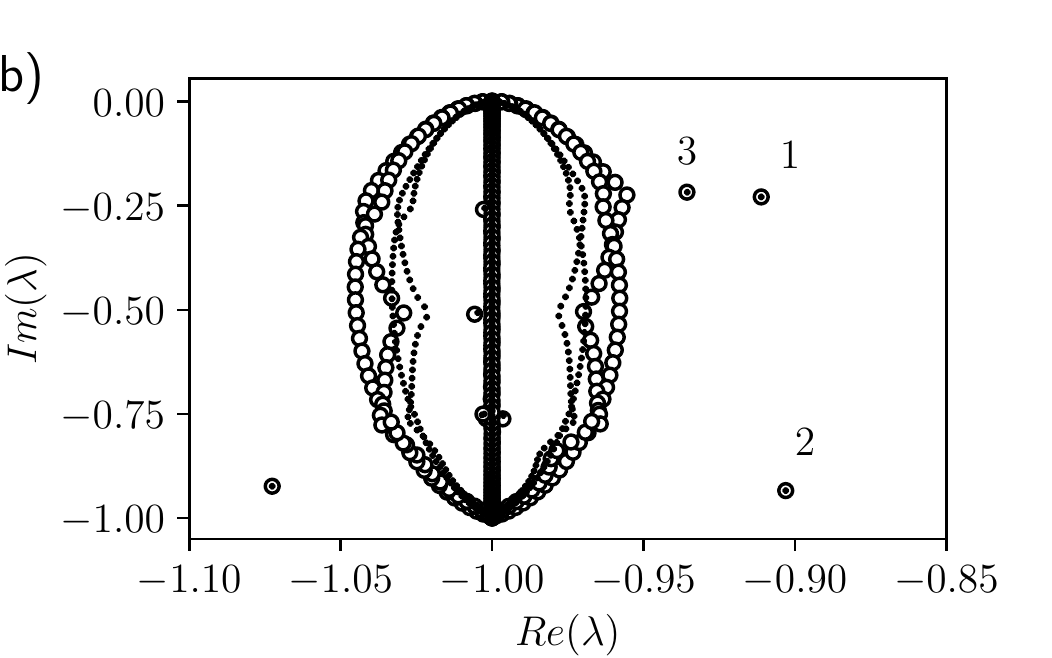}
\caption{a) Eigenvalue spectrum for $\Wi=1$, $\beta=0.05$, $k_x=1$, and $k_z=2$. b) Zoom-in on the most unstable part of the spectrum. The least stable eigenmodes, used in the non-linear analysis below, are denoted as "$1$", "$2$", and "$3$". Numerical convergence is insured by performing the linear stability analysis with $100$ (circles) and $150$ (dots) Chebyshev polynomials.}
\label{Fig:Spectra}
\end{figure}

\begin{figure}[htp]
\centering
\begin{tabular}{cc}
\includegraphics[width=0.48\textwidth]{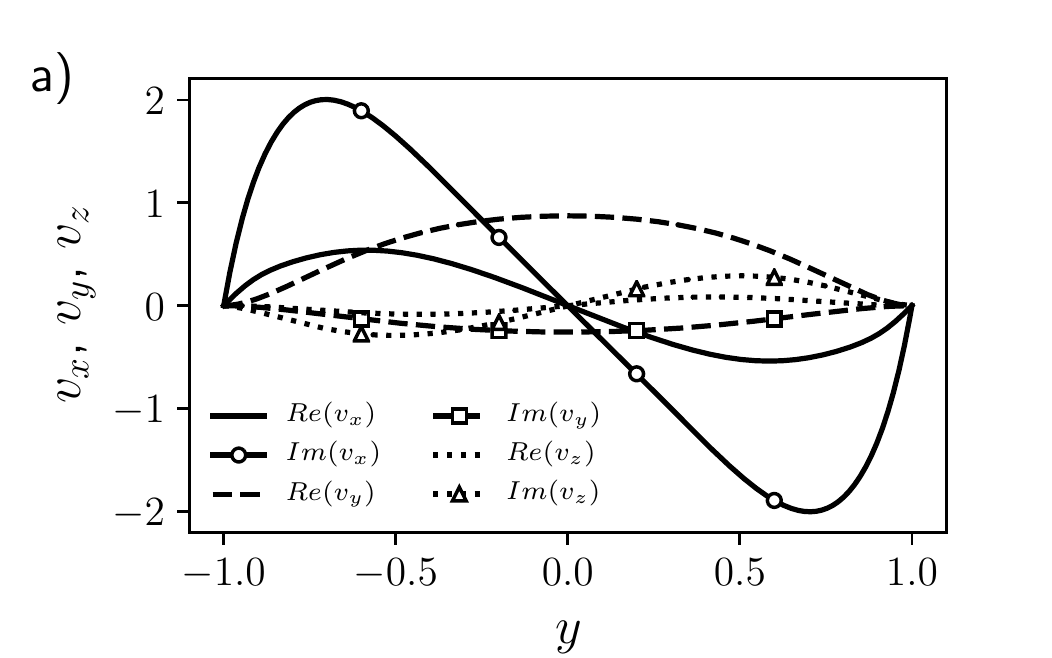} & 
\includegraphics[width=0.48\textwidth]{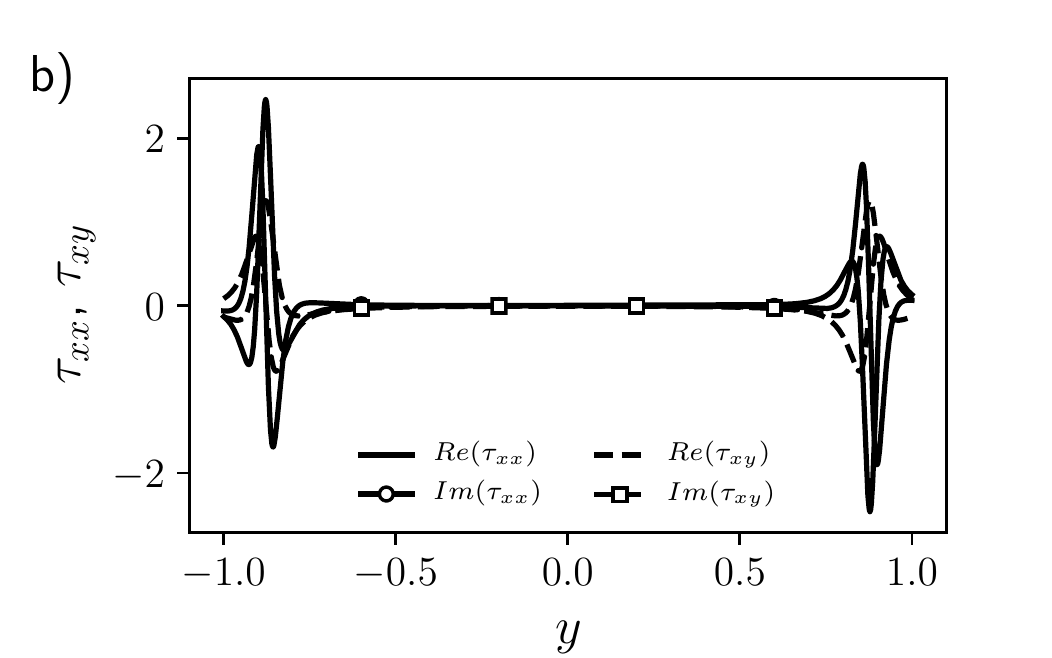} \\
\includegraphics[width=0.48\textwidth]{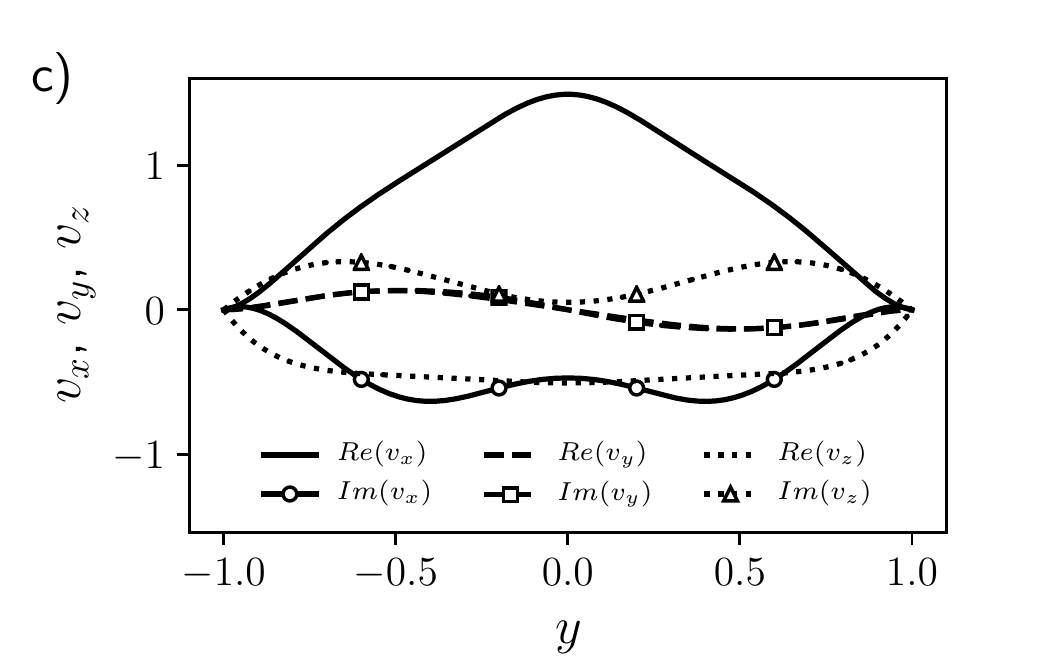} & 
\includegraphics[width=0.48\textwidth]{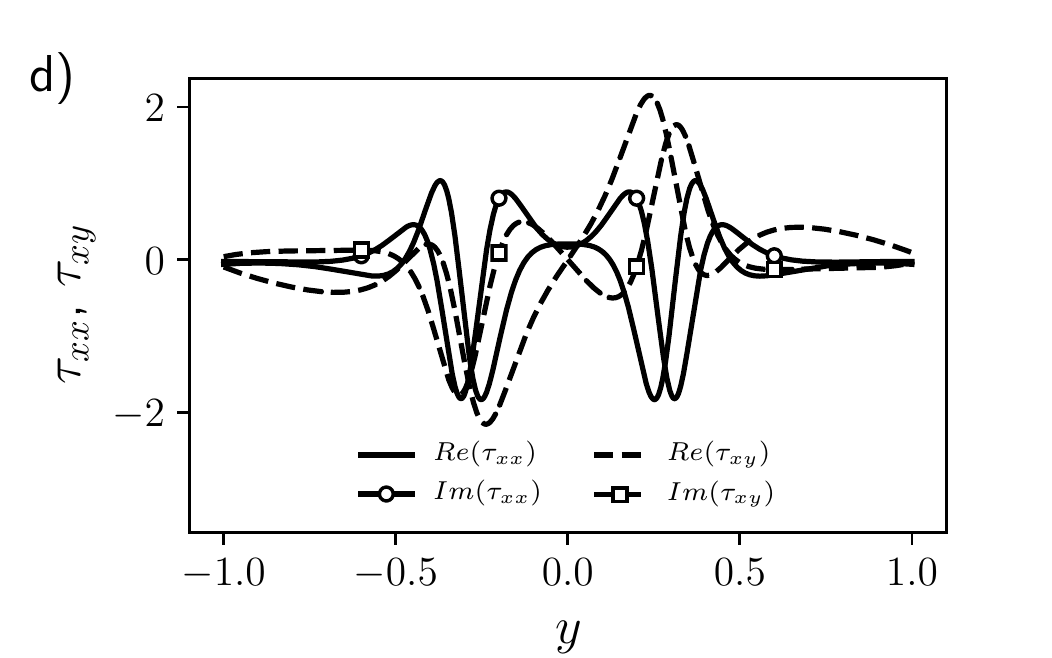} \\
\includegraphics[width=0.48\textwidth]{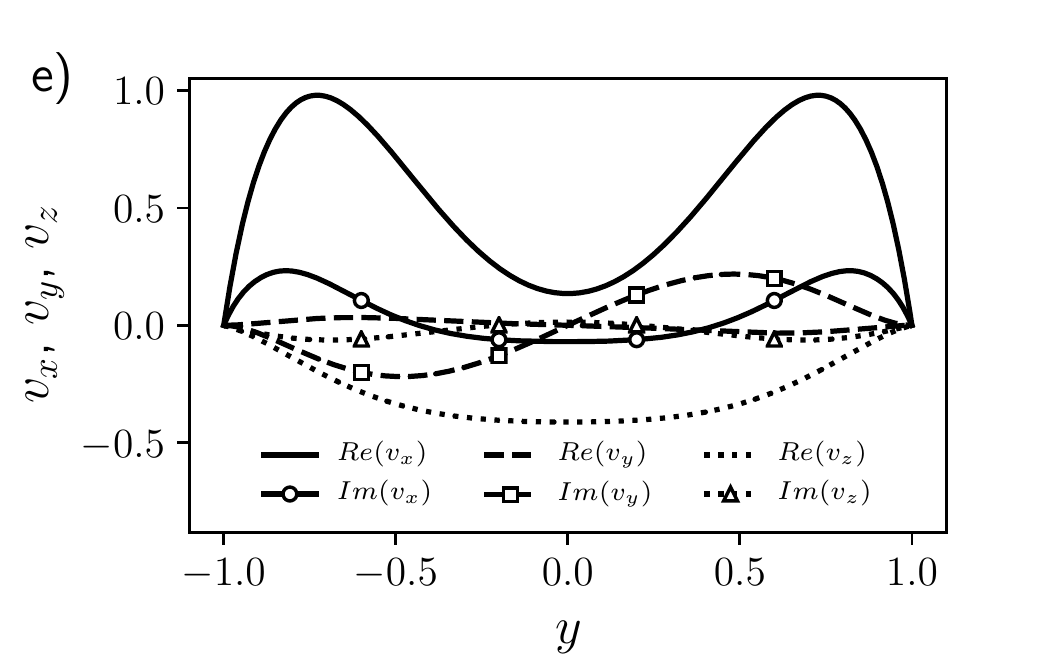} & 
\includegraphics[width=0.48\textwidth]{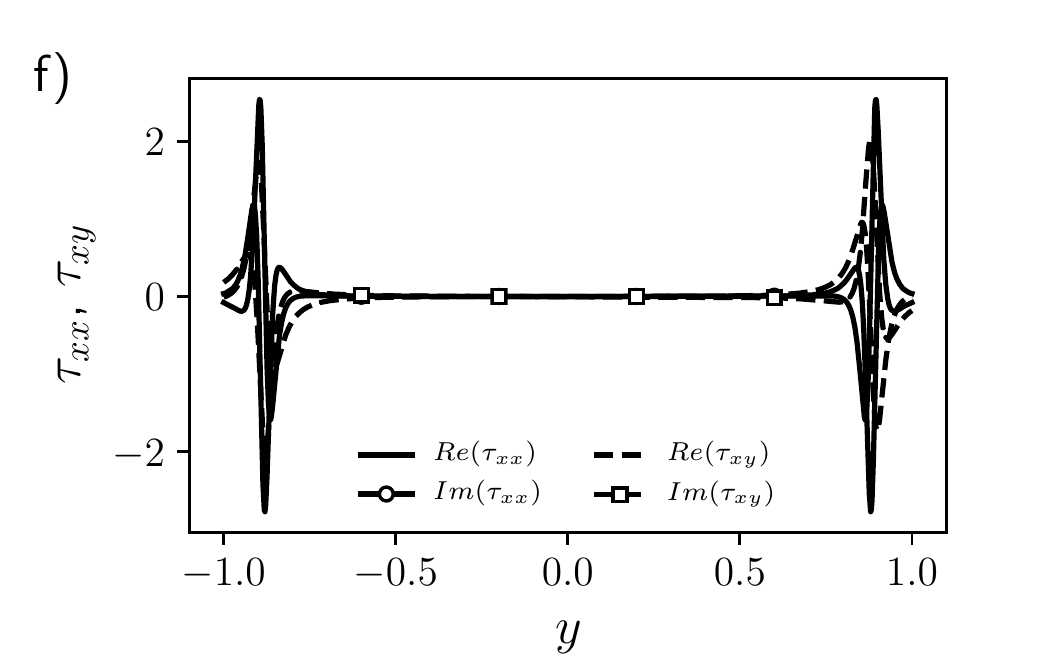}
\end{tabular}
\caption{Spatial profiles of the eigenmodes corresponding to the three right-most eigenvalues in Fig.\ref{Fig:Spectra}b). All eigenmodes are arbitrarily normalised. a) and b), c) and d), and e) and f) show the velocity and selected stress components corresponding to $\lambda_1$, $\lambda_2$, and $\lambda_3$, respectively.  For visualisation purposes, the velocities are scaled by the factors $10^4$, $10^2$, and $10^4$ in a), c) and e), respectively; $\tau_{xy}$ is scaled by the factors $20$, $10$, and $40$ in b), d) and f), respectively. In what follows, all spatial calculations are performed with $150$ Chebyshev polynomials.}
\label{Fig:Eigenmodes}
\end{figure}

In Fig.\ref{Fig:Spectra} we present an example of the eigenvalue spectrum, plotted as $Im(\lambda)$ vs $Re(\lambda)$. The general structure of the spectrum for $\Wi=1$, $\beta=0.05$, $k_x=1$, and $k_z=2$ is shown in Fig.\ref{Fig:Spectra}a), while Fig.\ref{Fig:Spectra}b) presents a zoom-in on the least stable part of the spectrum. As can be seen, all eigenvalues have $Re(\lambda)<0$, indicating linear stability. This is confirmed numerically for all values of $\Wi$, $\beta$, $k_x$ and $k_z$; see also Wilson \emph{et al.} \cite{Wilson1999}. The most prominent features in Fig.\ref{Fig:Spectra}, the balloon-like shapes at $Re(\lambda)=1/\Wi$ and $Re(\lambda)=1/(\beta \Wi)$, are numerical approximations to the continuous spectrum of the linear operator, which corresponds to the singular points of the linear problem \cite{Wilson1999}. The least unstable eigenvalues, which are used in the non-linear analysis below, are the right-most modes in Fig.\ref{Fig:Spectra}b), which we denote by $\lambda_1$, $\lambda_2$, and $\lambda_3$. (There is another discreet eigenvalue which is very close to the continuous spectrum, and we do not attempt to perform calculations for the associated eigenmode as it would require a very high spectral resolution.) The corresponding eigenmodes are presented in Fig.\ref{Fig:Eigenmodes}. As can be seen there, two of the eigenmodes are mostly pronounced close to the walls, and are related to the Gorodtsov-Leonov modes \cite{Gorodtsov1967} used in our previous work on plane Couette flow \cite{Morozov2005prl}, while the other one is mostly present in the middle of the geometry.

\begin{figure}[htp]
\centering
\begin{tabular}{cc}
\includegraphics[width=0.48\textwidth]{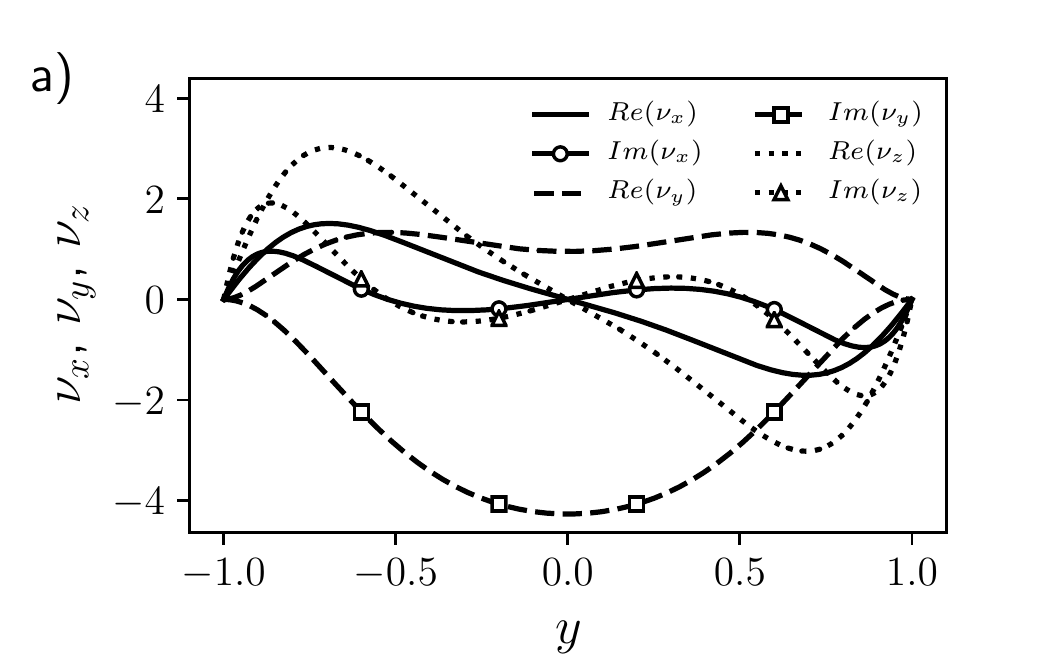} & 
\includegraphics[width=0.48\textwidth]{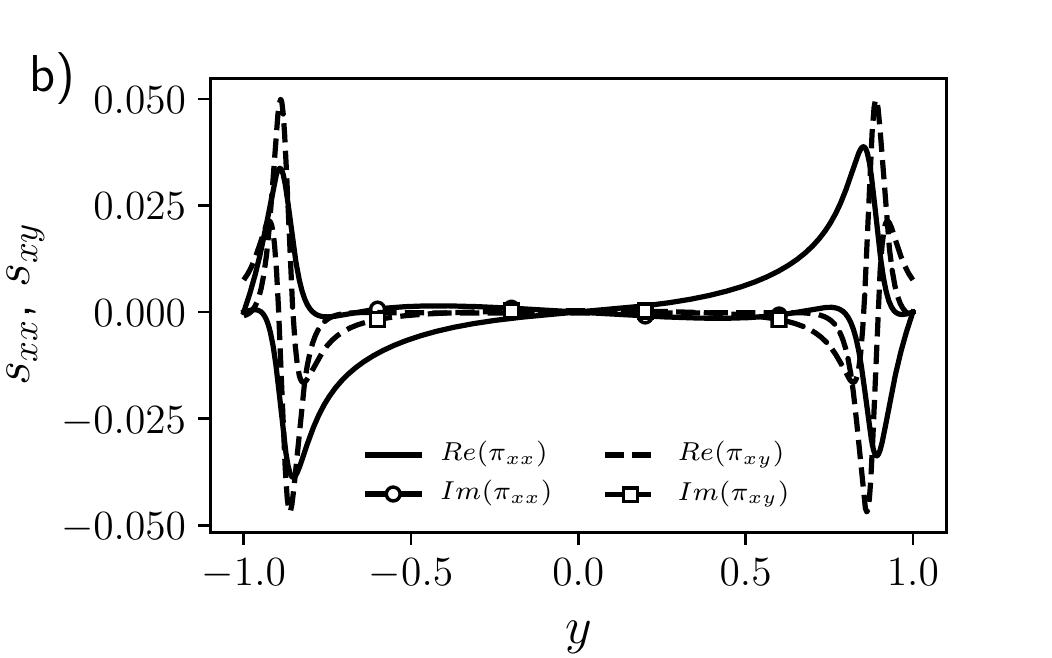}
\end{tabular}
\caption{Spatial profiles of the adjoint eigenmode (arbitrarily normalised) corresponding to $\lambda_1$ in Fig.\ref{Fig:Spectra}b). For visualisation purposes, the velocities and $s_{xx}$ are scaled by the factors $10^4$ and $30$, correspondingly.}
\label{Fig:Adjoint}
\end{figure}

Next, we calculate the eigenmodes of the adjoint problem, Eq.\eqref{adjoint_eigenfunctions}, with $\hat{\mathcal L}^\dagger$ defined in Appendix \ref{appendix_B}, by using the same numerical method, as above. As demonstrated in Section \ref{sect:AmplEq}, every eigenvalue of the linear problem $\lambda$ has its adjoint counterpart $\chi$, such that $\chi^* = \lambda$. Therefore, for the same parameters as above, the adjoint spectrum looks like Fig.\ref{Fig:Spectra}, with $Im(\chi) = -Im(\lambda)$. The adjoint eigenmodes, corresponding to the least stable eigenvalues $\lambda_1$, $\lambda_2$, and $\lambda_3$ share the same spatial features as their linear counterparts and are either confined to the walls or the bulk of the system; see Fig.\ref{Fig:Adjoint}, for example.

\begin{table}[htbp]
\centering
\begin{tabular}{r|cccccc}
$Wi$ & $Re(\lambda)$ & $Re(C_3)$ & $Re(C_5),\,\times 10^2$ & $Re(C_7),\,\times 10^3$ & $Re(C_9),\,\times 10^5$ & $Re(C_{11}),\,\times 10^7$ \\ 
\hline	 \\
2.0 & -0.4060 & 9.5015 & -1.3539 & -18.4446 & -12.2280 & -6.9523 \\ 
2.2 & -0.3612 & 9.2059 & -0.6473 & -14.3000 & -10.2052 & -4.9994 \\ 
2.4 & -0.3242 & 8.8739 & -0.0345 & -10.1102 & -8.1790 & -4.0287 \\ 
2.6 & -0.2934 & 8.4988 & 0.4652 & -6.1210 & -5.9845 & -3.1370 \\ 
2.8 & -0.2674 & 8.0893 & 0.8489 & -2.5689 & -3.7298 & -2.0734 \\ 
3.0 & -0.2452 & 7.6600 & 1.1256 & 0.4023 & -1.5917 & -0.8751 \\ 
3.2 & -0.2260 & 7.2250 & 1.3110 & 2.7499 & 0.2883 & 0.3327 \\ 
3.4 & -0.2094 & 6.7961 & 1.4228 & 4.5075 & 1.8361 & 1.4356 \\ 
3.6 & -0.1949 & 6.3819 & 1.4780 & 5.7515 & 3.0389 & 2.3649
\end{tabular}
\caption{Values of the non-linear coefficients $C$'s corresponding to $\lambda_1$ for $\beta=0.05$, $k_x=1$, and $k_z=2$, as functions of the Weissenberg number $\Wi$.}
\label{Table:R1}
\end{table}

\begin{table}[htbp]
\centering
\begin{tabular}{r|cccccc}
$Wi$ & $Re(\lambda)$ & $Re(C_3)$ & $Re(C_5),\,\times 10^2$ & $Re(C_7),\,\times 10^3$ & $Re(C_9),\,\times 10^5$ & $Re(C_{11}),\,\times 10^7$ \\ 
\hline	 \\
2.0 & -0.4193 & 14.1783 & -4.4338 & -74.3059 & -73.2408 & -82.2225 \\ 
2.2 & -0.3723 & 12.9427 & -2.5531 & -52.2014 & -49.8742 & -40.1729 \\ 
2.4 & -0.3335 & 11.9032 & -1.1161 & -35.3082 & -35.0532 & -24.7110 \\ 
2.6 & -0.3010 & 10.9746 & -0.0507 & -22.1785 & -24.0470 & -16.8609 \\ 
2.8 & -0.2734 & 10.1186 & 0.7063 & -12.1143 & -15.3407 & -11.2451 \\ 
3.0 & -0.2499 & 9.3208 & 1.2145 & -4.6404 & -8.4823 & -6.5830 \\ 
3.2 & -0.2297 & 8.5774 & 1.5293 & 0.6780 & -3.2760 & -2.7327 \\ 
3.4 & -0.2121 & 7.8888 & 1.6995 & 4.2694 & 0.4783 & 0.2575 \\ 
3.6 & -0.1969 & 7.2561 & 1.7664 & 6.5377 & 3.0289 & 2.4086
\end{tabular}
\caption{Values of the non-linear coefficients $C$'s corresponding to $\lambda_3$. The values of $\beta$, $k_x$ and $k_z$ are the same as in Table \ref{Table:R1}.}
\label{Table:R3}
\end{table}

In what follows, we use the three least stable modes as the starting point of the non-linear analysis presented in Section \ref{sect:AmplEq}, and assess whether they result in converged non-linear states. For each mode, we calculate the coefficients $C$'s as a function of the Weissenberg number $\Wi$, and use them to solve Eq.\eqref{threshold} for the amplitude $\Psi$ of the travelling-wave state. As explained  in Section \ref{sect:AmplEq}, we construct a series of solutions at progressively higher orders in the amplitude, see Eq.\eqref{Sms}, and study their convergence. We found that using the eigenmode associated with $\lambda_2$ does not lead to a converging series of amplitudes $\Psi$, while the other two eigenmodes lead to converging non-linear solutions, which we refer to as 'State 1' and 'State 3', and below we focus on these two modes. 

\begin{figure}[htp]
\centering
\begin{tabular}{cc}
\includegraphics[width=0.48\textwidth]{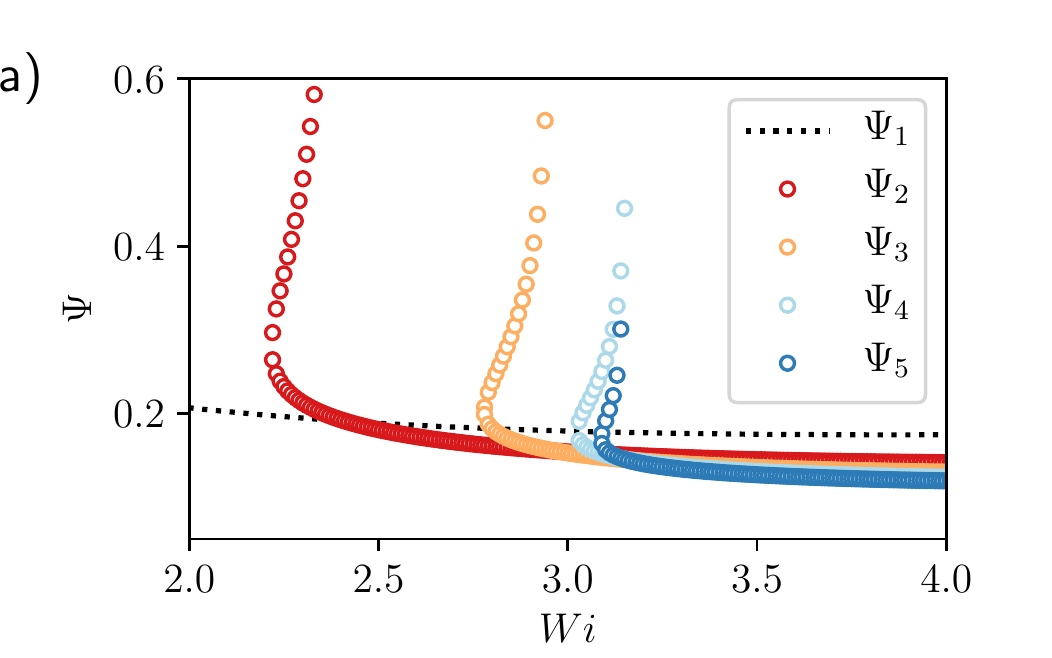} & 
\includegraphics[width=0.48\textwidth]{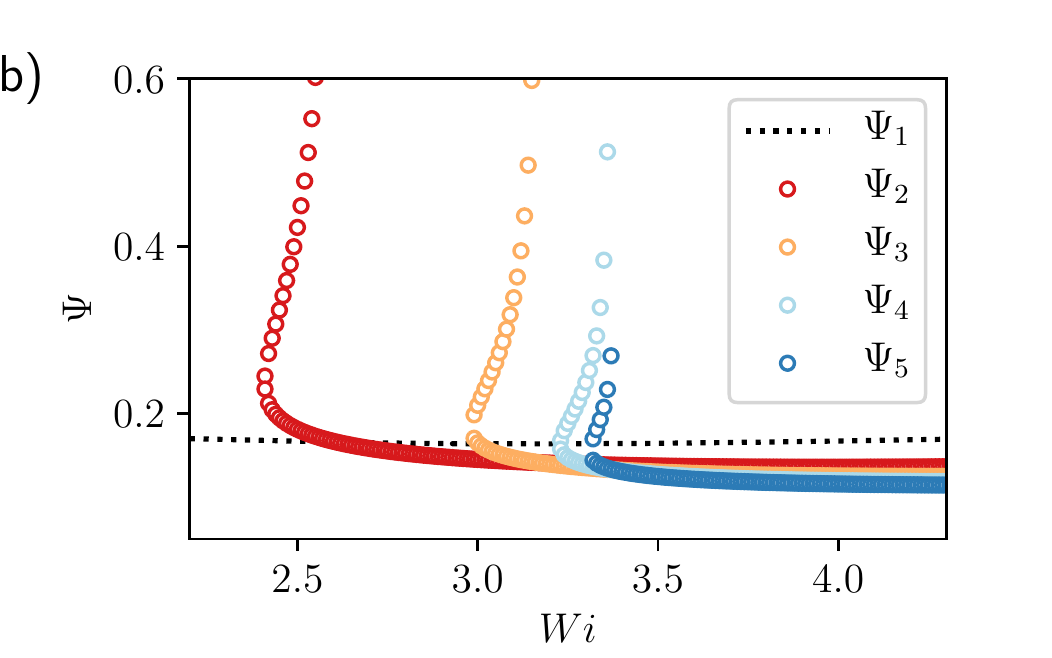}
\end{tabular}
\caption{Amplitudes $\Psi_m$ of the travelling wave solutions corresponding to a) $\lambda_1$ (State 1) and b) $\lambda_3$ (State 3) as functions of $\Wi$  for the case of plane Poiseuille flow; $m$ denotes the order $2m+1$ to which the expansion in powers of $\Phi$ in Eq. (\ref{amplitude_equation}) is taken. Note the similarity with Fig.1 for the case of plane Couette flow.
}
\label{Fig:bifurcation}
\end{figure}

\begin{figure}[htp]
\centering
\begin{tabular}{cc}
\includegraphics[width=0.65\textwidth]{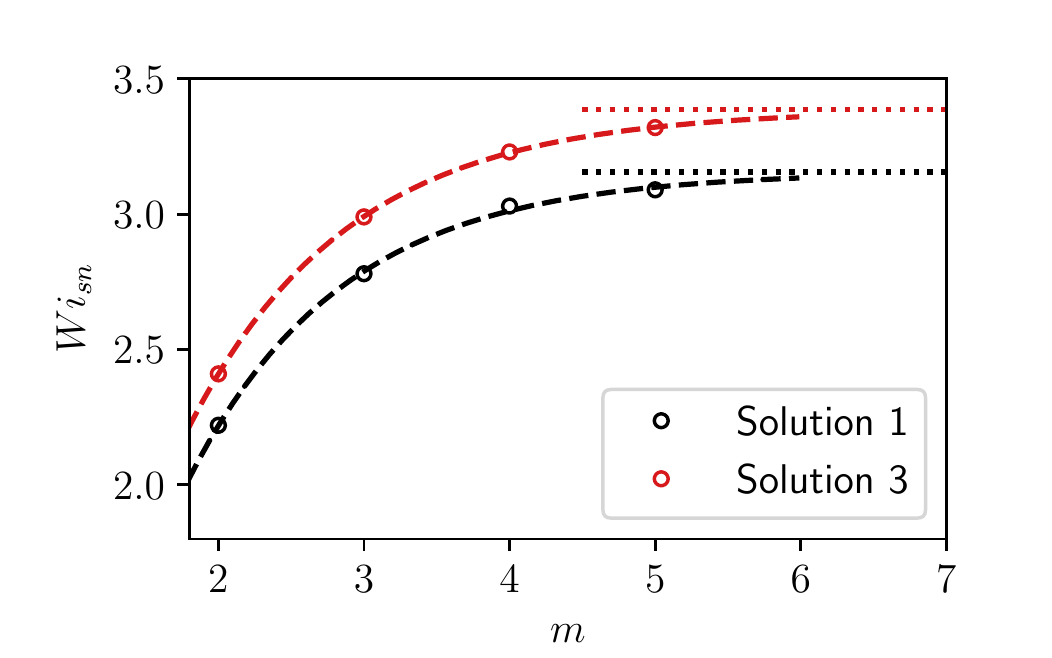}
\end{tabular}
\caption{Position of the saddle-nodes values $\Wi_{sn}$ as a function of the approximation order $m$, determined as the lowest value of $\Wi$ for which a real solution $\Psi_m$ exists. Circles are the values extracted from the data presented in Fig.\ref{Fig:bifurcation}. The dotted lines are exponential fits to the data given by $\Wi_{sn} = 3.15597 - 6.12468\exp(-0.938445m)$ (State 1), and $\Wi_{sn} = 3.38612 - 5.98429\exp(-0.906507m)$ (State 2).  The dotted lines denote thus extracted asymptotic values $\Wi^{(1)}_{sn}\approx3.16$ and $\Wi^{(3)}_{sn}\approx3.39$ for State 1 and 3, correspondingly.}
\label{Fig:saddle-node}
\end{figure}

The results of this procedure are presented in Fig.\ref{Fig:bifurcation}, while the values of the coefficients $C$'s for selected values of $\Wi$ are given in Tables \ref{Table:R1} and \ref{Table:R3}. For both states, the series of solutions $\Psi_m$ share the same features. Here, $m$ denotes the order $2m+1$ to which the expansion in powers of $\Phi$ in Eq. (\ref{amplitude_equation}) is taken.
For any $m>1$, the equation $S_m=0$ has no real solutions at small $\Wi$, two real solutions around the saddle-node bifurcation -- the value of $\Wi$ at which two solution branches appears for the first time, and one real solution for larger values of $\Wi$. 
The lower branches define the threshold amplitude required to destabilise the flow, while the upper branches are supposed to set the saturated value of the amplitude at the instability.
As can be seen from Fig.\ref{Fig:bifurcation}, the upper branches of all solutions seem to diverge rapidly close to the saddle-node value $\Wi_{sn}$, and the implications of this behaviour are discussed below.

\begin{figure}[htp]
\centering
\begin{tabular}{cc}
\includegraphics[width=0.48\textwidth]{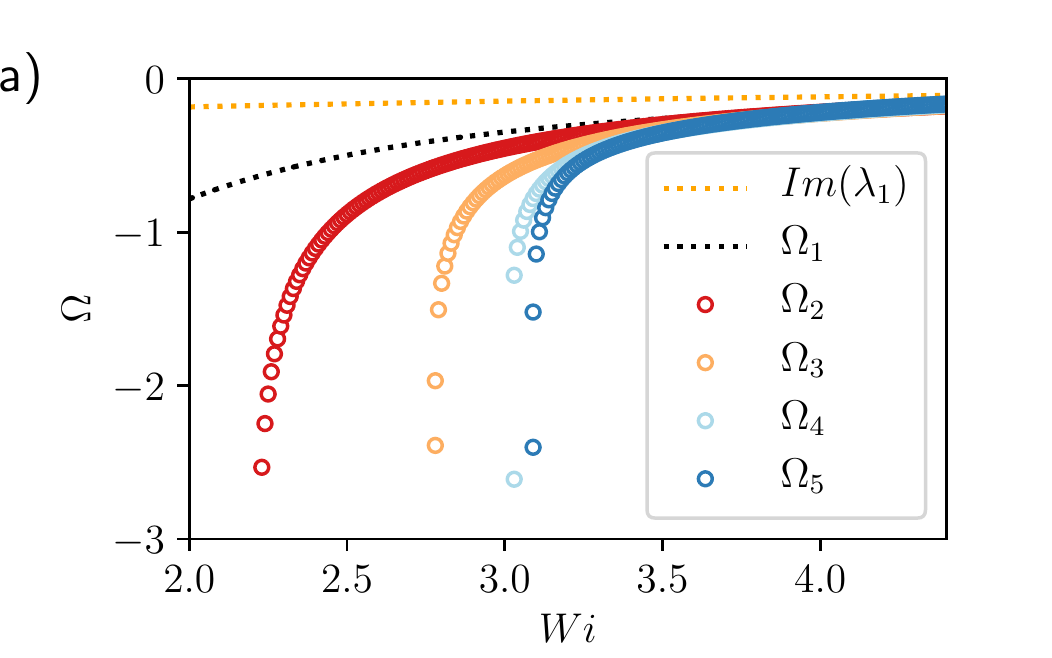} & 
\includegraphics[width=0.48\textwidth]{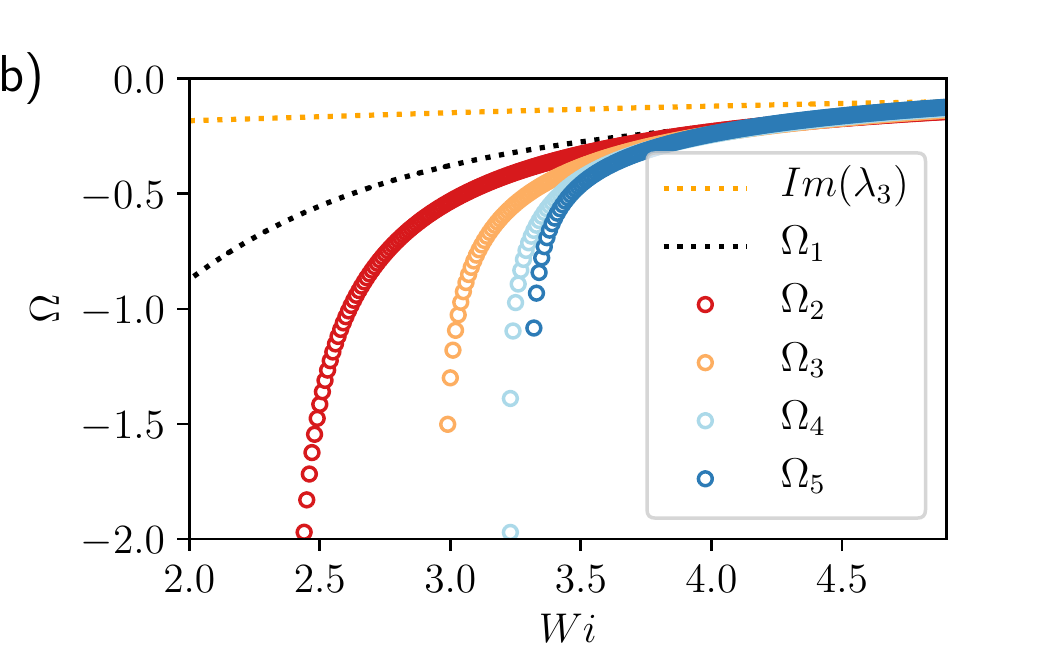}
\end{tabular}
\caption{The phase velocity $\Omega$ of the travelling-wave solution, Eq.\eqref{wavespeed}, calculated using $\Psi_m$ from Fig.\ref{Fig:bifurcation}, as a function of $\Wi$. a) State 1, b) State 3. }
\label{Fig:phases}
\end{figure}

For both states, we observe that the upper- and lower-branch values of $\Psi_m$ converge rapidly as $m$ is increased from $1$ to $5$. In Fig.\ref{Fig:saddle-node} we assess this convergence more quantitatively, by plotting the saddle-node values $\Wi_{sn}$ for each $\Psi_m$ as a function of $m$ (circles in Fig.\ref{Fig:saddle-node}). In the range $m=1\dots5$, corresponding to the amplitude equation expansion up to the eleventh order, the convergence is well-described by an exponential fit (dashed lines in Fig.\ref{Fig:saddle-node}), approaching $\Wi^{(1)}_{sn}\approx3.16$ and $\Wi^{(3)}_{sn}\approx3.39$ for States 1 and 3, correspondingly. The phase speed of the travelling-wave solutions, $\Omega$, calculated using the values of $\Psi_m$ presented above are shown in Fig.\ref{Fig:phases}. Again, we observe that the position of the saddle-node and the lower-branch values converge rapidly. Using the criterion presented in Section \ref{sect:AmplEq}, we conclude that our consecutive approximations to States 1 and 3 converge towards physical solutions that we now examine in more detail.

\begin{figure}[htp]
\centering
\includegraphics[width=0.95\textwidth]{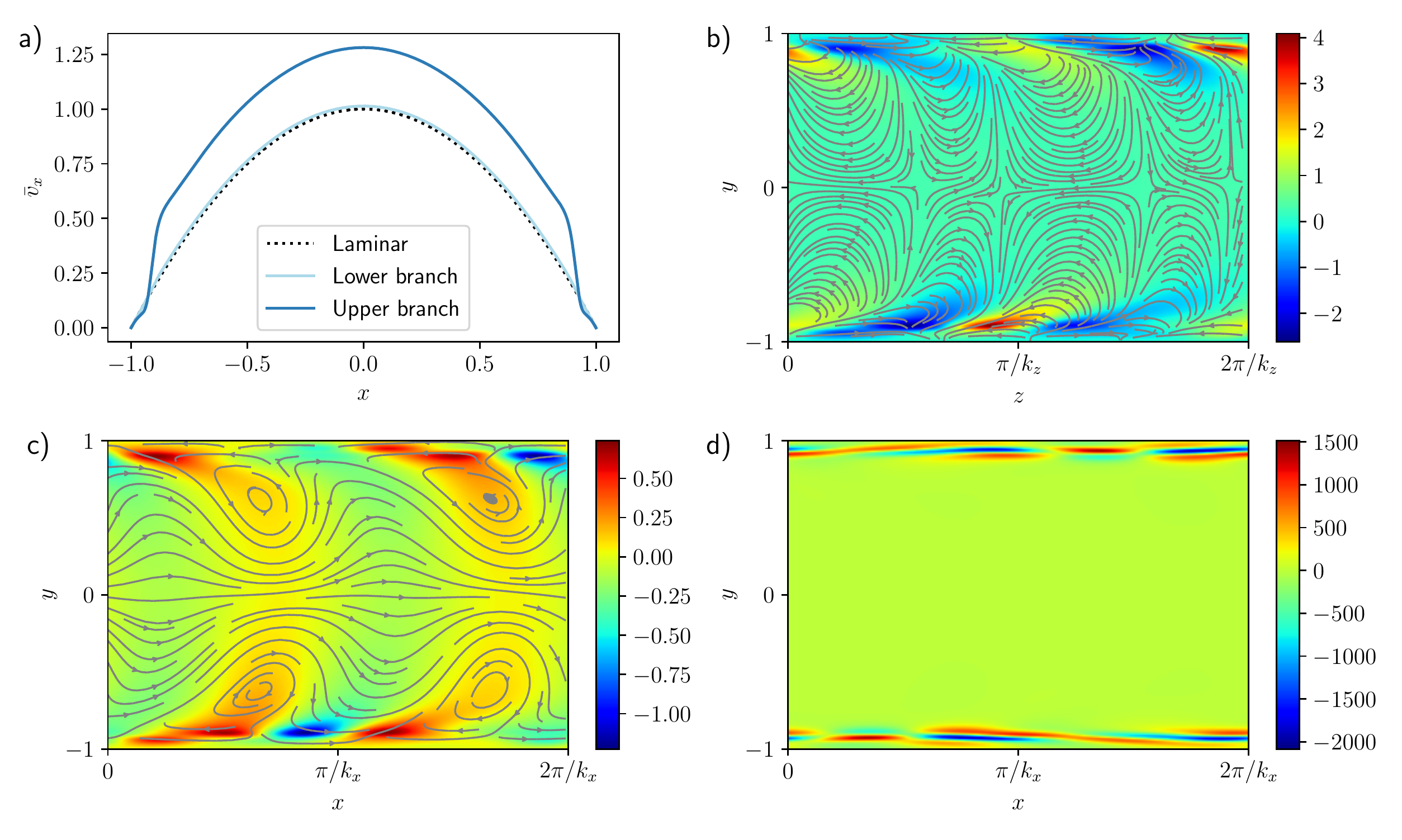}
\caption{Spatial profiles corresponding to State 1 at $\Wi=3.12$ and $\Psi=0.22$ (the upper-branch solution). a) The mean streamwise velocity profile (the lower-branch solution has $\Psi=0.15$). b) The velocity field in the $yz$-plane at $x=0$: streamlines show the $v_y$ and $v_z$ velocities, the colour shows the $v_x-\mathcal{U}_0(y)$ profile. c) The velocity field in the $xy$-plane at $z=0$:  streamlines show the $v_x-\mathcal{U}_0(y)$ and $v_y$ velocities, the colour shows the $v_z$ profile. d) Deviation from the laminar profile of the normal stress component $\tau_{xx}$ in the $xy$-plane at $z=0$.}
\label{Fig:SpatialState1}
\end{figure}

To gain insight into the spatial structure of the travelling-wave solutions reported above, we now plot the mean velocity profile $\bar{v}_x$ for State 1, calculated by averaging $v_x$ over $x$ and $z$. 
In Fig.\ref{Fig:SpatialState1}a) we present the mean profile for $\Wi=3.12$, which is close enough to the saddle-node point for $m=5$ so that $\Psi_5$ has two values,  $0.15$ and $0.22$, corresponding to the lower- and upper-branches, respectively. While the lower-branch mean velocity profile is quite close to the laminar profile $\mathcal{U}_0(y) =1-y^2$, the upper-branch profile looks very different, with sharp velocity gradients close the walls and a shifted parabolic-like profile in the bulk of the system. Surprisingly, the centre-line velocity appears to be \emph{larger} than its laminar counterpart. In Fig.\ref{Fig:SpatialState1}b) we present the velocity profile at $x=0$ in the $yz$-plane, which is perpendicular to the streamwise direction: the in-plane components $v_y$ and $v_z$ are shown as arrows, while the streamwise velocity $v_x-\mathcal{U}_0(y)$ is given by the colour. One can clearly see two arrays of streamwise vortices next to each wall superimposed onto the corresponding arrays of high- and low-velocity streamwise streaks. 
Fig.\ref{Fig:SpatialState1}c) shows a similar velocity profile at $z=0$ in the $xy$-plane, where arrows now trace the in-plane components $v_x-\mathcal{U}_0(y)$ and $v_y$, while the colour gives the spatial profile of the spanwise velocity $v_z$. Finally, in Fig.\ref{Fig:SpatialState1}d) we plot the largest component of the polymeric stress tensor, $\tau_{xx}$, which is a deviation from the laminar stress, in the $xy$-plane. Most of the stress is concentrated close to the boundaries consistent with the presence of sharp velocity gradients there. 

The profiles presented in Fig.\ref{Fig:SpatialState1}b)-d) correspond to the upper branch of State 1. The lower branch profiles have a very similar structure, albeit with a significantly smaller amplitude, and are not presented here. In a similar fashion, the mean profiles and the spatial structure of State 3 bear strong similarities with State 1, and are, therefore, also omitted here.

Finally, the two states presented here are shown for the somewhat randomly selected values of the wavenumbers $k_x$ and $k_z$. Preliminary studies show that for small values of $\beta$, converged solutions similar to States 1 and 3 persist for a wide range of wavenumbers, provided both $k_x$ and $k_z$ are not too big (typically, smaller than $k_x\sim k_z\sim 3$. For larger values of $\beta$, this region shrinks, which is either related to the convergence properties of our technique, or is connected to the actual shrinking of the region of existence of such travelling-wave solutions. This point is deferred to future studies.

\section{Discussion and Conclusion}
\label{sect:discussion}

Results presented in this work further corroborate our previous claim \cite{Meulenbroek2003,Bertola2003,Morozov2005prl,Morozov2007} that while parallel shear flows of viscoelastic fluids are linear stable, they exhibit sub-critical instabilities that lead directly to a chaotic state.  Early experiments by Bonn \emph{et al.} \cite{Bonn2011} and in particular the more recent detailed and systematic experiments  by Arratia \emph{et al.} \cite{Pan2013,Qin2017} confirm the existence of such sub-critical instabilities in channel flows of dilute polymer solutions and demonstrates that the chaotic state observed there is related to the phenomenon of purely elastic turbulence previously only reported in shear flows with curved streamlines \cite{Groisman2000,Groisman2004,Burghelea2007}.

The emergent scenario of the transition in parallel shear flows of viscoelastic fluids parallels that for their Newtonian counterparts. Recently, significant progress was made in understanding the transition to turbulence in pipe, plane Couette and channel flows of Newtonian fluids by studying it from the dynamical systems' point of view \cite{Eckhardt2008,barkley2016}. The key ingredients there are the so-called \emph{coherent structures}, the exact solutions of the Navier-Stokes equations, either travelling waves or periodic orbits, that are three-dimensional but relatively simple. These solutions appear through a sub-critical bifurcation
and correspond to saddles in the phase space of the flow: while their upper and lower branches are linearly unstable, their vicinity is attractive; also, their number increases with the Reynolds number. When the phase space is sufficiently populated with such solutions, a turbulent trajectory performing a random walk between a large number of saddle-like states gets trapped for a very long time. Coupled to the phenomenon of splitting of localised exact coherent states, this scenario firmly places the transition to Newtonian turbulence within the directed percolation universality class.

Our results indicate that a similar scenario might also be at work in the viscoelastic case. The solutions presented here and in our previous work \cite{Morozov2005prl} might form the phase space scaffold of the purely elastic turbulence, and verifying their existence should be paramount to understanding its mechanism. The amplitude-equation type technique employed here attempts to construct a non-linear solution as an asymptotic series, and, as such, it has a limited radius of convergence. While the lower branches of our solutions, indicative of the amplitude threshold required to trigger the turbulent state, are well-converged, the upper branches disappear in a close vicinity of the saddle-node bifurcation. This is either related to the radius of convergence of the asymptotic series, Eq.\eqref{amplitude_equation}, as mentioned above, or can be the direct consequence of the upper-branch non-linear state being turbulent, and the failure of the technique to capture it.

The solutions that we found in this work appear at the values of the Weissenberg number, $\Wi_{sn}\sim 3$, that are somewhat lower than the onset of turbulence values reported in experiments in channel, $\Wi_{onset}\sim 5$\cite{Pan2013,Qin2017}, and pipe flows, $\Wi_{onset}\sim 4$ \cite{Bertola2003,Bonn2011}. This is not surprising since the fluids used in those works had significantly higher values of $\beta$ than $0.05$ used in this work, and, moreover, exhibited various degrees of shear thinning, which is not included in the present analysis. Both factors would result in higher values of $\Wi_{sn}$ than presented here. Also, the dynamical systems scenario of the transition, if applicable in the viscoelastic case, would imply that $\Wi_{sn}$ should be smaller than the value of $\Wi_{onset}$ at which sustained turbulence can be observed, providing yet another explanation for the difference.\footnote{One should also keep in mind that we study a mode which is fully periodic in the streamwise direction, whereas in the experiments the flow is perturbed by a small number of cylindrical obstacles in the flow channel.} We do not argue in favour of either of the explanations and, instead, draw a more conservative conclusion that our work provides further evidence for a subcritical (nonlinear) instability scenario at moderate values of the Weissenberg number, and that it demonstrates that exact coherent solutions do exist in this type of viscoelastic flows and proper numerical investigation of such states is required.

While experiments provide very limited information about the spatial structure of purely elastic turbulence in parallel shear flows \cite{Bertola2003,Bonn2011,Pan2013,Qin2017}, our work sheds light on what profiles might be expected in such geometries. First, we observe that, according to our calculations, the turbulent mean velocity profiles should exhibit a larger centre-line velocity than their laminar counterparts, see Fig.\ref{Fig:SpatialState1}a). This is in stark contrast with the Newtonian turbulent mean profiles, which are plug-like due to the momentum re-distribution between the walls and the bulk, and are always slower than the corresponding laminar velocity field in the middle of the gap. Intriguingly, a similar profile was reported under certain conditions in recent experiments on pipe and channel flows of rather concentrated polymer solutions \cite{Poole2016}, where it was attributed to shear-thinning. The profiles reported here suggest that it might be a more generic feature related to the elasticity of the fluid.  
Our second observation comprises the existence of streamwise vortices and streaks in a plane perpendicular to the flow direction, Fig.\ref{Fig:SpatialState1}b). These structures are a hallmark of Newtonian coherent structures, and it appears that they also play a role in viscoelastic non-linear solutions. This is, perhaps, not surprising since they feature prominently in non-normal growth analysis by Kumar and Jovanovi\'{c} \cite{Jovanovic2010,Jovanovic2011}. The associated stresses (not shown) are in line with the positions of large velocity gradients in-between the streamwise vortices, although both structures are tilted due to the travelling-wave nature of whole the state.
Finally, the plane perpendicular to the spanwise $z$-direction contains widely-spaced co-rotating vortices interlaid with expanding-contracting streamlines and large, wall-localised stresses. Although such structures have not been reported in the literature, they are consistent with the near-wall mixing patterns observed by Qin \emph{et al.} \cite{Qin2017}.

Although we are confident that we obtained converged non-linear states, their existence needs to be verified numerically, by searching for steady-state, travelling-waves, and periodic orbits, using a Newton-Raphson-type algorithm. Until recently, such calculations were not feasible as a Newton-Raphson step is akin to a time-iteration step for the same equations, and viscoelastic constitutive models are notoriously difficult to time-step at sufficiently high Weissenberg numbers (the so-called High-Weissenberg Number Problem \cite{OwensPhillips}). In the past years, there emerged a class of numerical techniques to ensure positive-definiteness of the conformation tensor (absence thereof was implicated as a cause of the High-
Weissenberg Number Problem), led by the log-conformation algorithm \cite{Fattal2005,Pimenta2017}, and a combination of such techniques with a Newton-Raphson-type algorithm should be able to overcome this problem. The states predicted in this work can then serve as a good initial guess for the search algorithm. Once found, upper branches of these solutions should be studied for their linear stability that will assess whether the transition mechanism in viscoelastic fluids bears similarities with its Newtonian counterpart. This work will be a subject of our future studies.

\newpage

\begin{appendix}
\section{Explicit form of the matrix equation}
\label{appendix_A}
As discussed in the main text, the equations of motion Eqs.\eqref{Stokes}-\eqref{boundary_conditions} can be written in the matrix form
\begin{equation}
\label{app_nle}
\hat{\mathcal L}V + \hat{A}\frac{\partial V}{\partial t} = N\left(V,V\right),
\end{equation}
where $V=\left(v_x,v_y,v_z,\tau_{xx},\tau_{xy},\tau_{xz},\tau_{yy},\tau_{yz},\tau_{zz},p\right)^\dagger$ is the dimensionless deviation of the hydrodynamic
fields from their laminar values (we have dropped the superscript ''1'' in Eqs.\eqref{stresspertub} and \eqref{velpertub} to simplify notation). The operator $\hat{A}$ is given by a constant diagonal matrix
\begin{align}
\Wi\,\,\text{diag}\Big(0,0,0,0,1,1,1,1,1,1 \Big),
\label{matrix_A}
\end{align}
while the linear operator $\hat{\mathcal L}$ is define by it action on $V$
\begin{eqnarray}
&&\hat{\mathcal{L}} V= \nonumber \\
&& \qquad\nonumber \\
&&\qquad\begin{pmatrix}
-\partial_x p + \beta\nabla^2 v_x  + (1-\beta)\Big[ \partial_x \tau_{xx}   + \partial_y\tau_{xy} + \partial_z\tau_{xz} \Big] \\
-\partial_y p + \beta\nabla^2 v_y  + (1-\beta)\Big[\partial_x \tau_{xy}   +  \partial_y\tau_{yy} + \partial_z\tau_{yz}\Big] \\
-\partial_z p + \beta\nabla^2 v_z  + (1-\beta)\Big[\partial_x \tau_{xz}   + \partial_y\tau_{yz} + \partial_z\tau_{zz} \Big]\\
\partial_x v_x +  \partial_y v_y + \partial_z v_z \\

\hat{X}_1 \tau_{xx} - 2\,\Wi\,\mathcal{U}_0'\tau_{xy} - 2  \hat{X}_2 v_x + 4\,\Wi^2\,\mathcal{U}_0'\,\mathcal{U}_0'' v_y \\

\hat{X}_1 \tau_{xy} - \Wi\,\mathcal{U}_0'\tau_{yy} -\partial_y v_x + \left( \Wi\,\mathcal{U}_0'' -  2\,\Wi^2\,\mathcal{U}_0'^2 \partial_x - \partial_x \right) v_y + \Wi\,\mathcal{U}_0' \partial_z v_z \\

\hat{X}_1 \tau_{xz} - \Wi\,\mathcal{U}_0'\tau_{yz} - \partial_z v_x - \hat{X}_2 v_z \\

\hat{X}_1 \tau_{yy} - 2\left( \partial_y + \Wi\,\mathcal{U}_0'\,\partial_x \right) v_y \\

\hat{X}_1 \tau_{yz} - \partial_z v_y - \left( \partial_y + \Wi\,\mathcal{U}_0'\,\partial_x \right) v_z \\

\hat{X}_1 \tau_{zz} - 2\partial_z v_z 
\end{pmatrix}.
\end{eqnarray}
Here, $\hat{X}_1 = 1+ \Wi\,\mathcal{U}_0\,\partial_x$ and $\hat{X}_2 = \left(1+2\,\Wi^2\,\mathcal{U}_0'^2 \right)\partial_x + \Wi\,\mathcal{U}_0'\,\partial_y$, and the dimensionless laminar velocity profile is given by $\mathcal{U}_0(y) = 1-y^2$.

The r.h.s. of Eq.\eqref{app_nle} represents the non-linear terms in the original equations Eqs.\eqref{Stokes}-\eqref{boundary_conditions}, and is given by a bilinear form
\begin{eqnarray}
&&N\left(V^{(A)},V^{(B)}\right) = - \left(\bm{v}^{(A)}\cdot{\mathbf \nabla}\right) \hat{A}\, V^{(B)} \nonumber \\
&& +\, \Wi\begin{pmatrix}
0 \\
0 \\
0 \\
0 \\
\\
2 \left[ \tau^{(A)}_{xx} \partial_x v^{(B)}_x + \tau^{(A)}_{xy} \partial_y v^{(B)}_x + \tau^{(A)}_{xz} \partial_z v^{(B)}_x \right] \\
\\
\tau^{(A)}_{xx} \partial_x v^{(B)}_y - \tau^{(A)}_{xy} \partial_z v^{(B)}_z + \tau^{(A)}_{xz} \partial_z v^{(B)}_y + \tau^{(A)}_{yy} \partial_y v^{(B)}_x 
+ \tau^{(A)}_{yz} \partial_z v^{(B)}_x \\ 
\\
\tau^{(A)}_{xx} \partial_x v^{(B)}_z + \tau^{(A)}_{xy} \partial_y v^{(B)}_z - \tau^{(A)}_{xz} \partial_y v^{(B)}_y + \tau^{(A)}_{yz} \partial_y v^{(B)}_x 
+ \tau^{(A)}_{zz} \partial_z v^{(B)}_x  \\
\\
2 \left[ \tau^{(A)}_{xy} \partial_x v^{(B)}_y + \tau^{(A)}_{yy} \partial_y v^{(B)}_y + \tau^{(A)}_{yz} \partial_z v^{(B)}_y \right] \\
\\
\tau^{(A)}_{xy} \partial_x v^{(B)}_z + \tau^{(A)}_{xz} \partial_x v^{(B)}_y + \tau^{(A)}_{yy} \partial_y v^{(B)}_z - \tau^{(A)}_{yz} \partial_x v^{(B)}_x 
+ \tau^{(A)}_{zz} \partial_z v^{(B)}_y  \\
\\
2 \left[ \tau^{(A)}_{xz} \partial_x v^{(B)}_z + \tau^{(A)}_{yz} \partial_y v^{(B)}_z + \tau^{(A)}_{zz} \partial_z v^{(B)}_z \right]

\end{pmatrix}. \nonumber
\end{eqnarray}
Obviously, $N\left(A,B\right) \neq N\left(B,A\right)$.

\section{Adjoint operator}
\label{appendix_B}

The actual form of the adjoint operator $\hat{\mathcal L}^\dagger$ is obtained by using in Eq.\eqref{definition_adjoint} the expression for the linear operator $\hat{\mathcal L}$, defined above, and performing integration-by-parts. Again, we define $\hat{\mathcal L}^\dagger$ by its action on a mode 
$W=\left(\nu_x,\nu_y,\nu_z,s_{xx},s_{xy},s_{xz},s_{yy},s_{yz},s_{zz},\pi\right)^\dagger$, which is given by
\begin{eqnarray}
&&\hat{\mathcal L}^\dagger W= \nonumber \\
&& \qquad\nonumber \\
&&\qquad\begin{pmatrix}
-\partial_x \pi +\beta \nabla^2 \nu_x + 2 \left( \partial_x +2\,\Wi^2\,\mathcal{U}_0'^2 \partial_x + \Wi\,\mathcal{U}_0'' + \Wi\,\mathcal{U}_0'\partial_y\right) s_{xx}\\
+ \partial_y s_{xy}  + \partial_z s_{xz} \\
\\
-\partial_y \pi +\beta \nabla^2 \nu_y + 4\,\Wi^2\,\mathcal{U}_0'\,\mathcal{U}_0'' s_{xx} + \left( \partial_x +2\,\Wi^2\,\mathcal{U}_0'^2 \partial_x + Wi\,\mathcal{U}_0'' \right) s_{xy} \\ 
+ 2 \left( \partial_y + \Wi\,\mathcal{U}_0'\partial_x \right) s_{yy} + \partial_z s_{yz} \\
\\
-\partial_z \pi +\beta \nabla^2 \nu_z - \Wi\,\mathcal{U}_0'\partial_z s_{xy} + \left( \partial_x +2\,\Wi^2\,\mathcal{U}_0'^2 \partial_x \right) s_{xz}\\ + \Wi \left( \mathcal{U}_0'' + \mathcal{U}_0'\partial_y \right) s_{xz}
 + \left( \partial_y + \Wi\,\mathcal{U}_0'\partial_x \right) s_{yz} + 2 \partial_z s_{zz} \\
\\
\partial_x \nu_x +  \partial_y \nu_y + \partial_z \nu_z \\
\\
\hat{Y}_1 s_{xx} - (1-\beta)\partial_x \nu_x \\
\\
\hat{Y}_1 s_{xy} - 2\,\Wi\,\mathcal{U}_0's_{xx} - (1-\beta)\left( \partial_y \nu_x + \partial_x \nu_y \right) \\
\\
\hat{Y}_1 s_{xz} - (1-\beta)\left( \partial_x \nu_z + \partial_z \nu_x \right) \\
\\
\hat{Y}_1 s_{yy} - \Wi\,\mathcal{U}_0's_{xy} - (1-\beta)\partial_y \nu_y \\
\\
\hat{Y}_1 s_{yz} - \Wi\,\mathcal{U}_0's_{xz} - (1-\beta)\left( \partial_y \nu_z + \partial_z \nu_y \right) \\
\\
\hat{Y}_1 s_{zz} - (1-\beta)\partial_z \nu_z
\end{pmatrix},
\end{eqnarray}
where $\hat{Y}_1 = 1- \Wi\,\mathcal{U}_0\,\partial_x$. The adjoint velocities $\nu$ are subject to the boundary conditions
\begin{align}
\nu_x=\nu_y=\nu_z=0\,,\qquad y=\pm 1,
\end{align}
which follow from the requirement that the boundary terms from the integration by parts in Eq.\eqref{definition_adjoint} vanish for any $V_1$ satisfying  Eq.\eqref{boundary_conditions}.

\section{Higher-order coefficients for the amplitude equation}
\label{appendix_C}

In Section \ref{sect:AmplEq} we demonstrated how the coefficient $C_3$ can be determined from Eq.\eqref{substituted}. Here, we list the expressions for higher-order coefficients $C$'s and the associated functions $u_n^{(m)}(y)$ from Eq.\eqref{Un}, obtained in the manner discussed in Section \ref{sect:AmplEq}.

To $O\left(\Phi |\Phi|^4\right)$, we have
\begin{align}
\hat{\mathcal{L}}\left( e^{3 i \xi} u_3^{(3)}\right) + 3\lambda  \hat{A} e^{3 i \xi} u_3^{(3)} =  \bar{N}\left( e^{i \xi}V_0,e^{2 i \xi} u_2^{(2)}\right),
\end{align}
\begin{align}
 \hat{\mathcal{L}}\left( e^{2 i \xi} u_2^{(4)}\right) &+ \left(3\lambda+\lambda^*\right)  \hat{A} e^{2 i \xi} u_2^{(4)} \nonumber \\
&  = \bar{N}\left( e^{-i \xi}V_0^*,e^{3 i \xi} u_3^{(3)}\right) +  \bar{N}\left( u_0^{(2)} ,e^{2 i \xi} u_2^{(2)}\right)
- 2 C_3 e^{2 i \xi} u_2^{(2)},
\end{align}
\begin{align}
\hat{\mathcal L} u_0^{(4)} + 2\left(\lambda + \lambda^* \right)\hat{A} u_0^{(4)} = 
N\left(u_0^{(2)},u_0^{(2)}\right)
+ \bar{N}\left(e^{2i\xi}u_2^{(2)}, e^{-2i\xi}u_2^{(2)*} \right)
-\left( C_3 +C_3^*\right) u_0^{(2)},
\end{align}
which yields
\begin{align}
& C_5 = \frac{1}{\Delta}\biggl\langle e^{i\xi} W_0 \biggl\vert 
   \bar{N}\left(e^{i\xi} V_0, u_0^{(4)}\right) 
+ \bar{N}\left(e^{-i\xi} V_0^*,e^{2 i\xi} u_2^{(4)}\right)
+ \bar{N}\left(e^{-2i\xi} u_2^{(2)*},e^{3 i\xi} u_3^{(3)}\right)
\biggr\rangle,
\end{align}
where, as in Section \ref{sect:AmplEq},
\begin{align}
\Delta = \langle e^{i\xi} W_0(y) \vert e^{i\xi} \hat{A} V_0(y)\rangle.
\end{align}

To $O\left(\Phi |\Phi|^6\right)$, we have
\begin{align}
\hat{\mathcal{L}}\left( e^{4 i \xi} u_4^{(4)}\right) + 4\lambda  \hat{A} e^{4 i \xi} u_4^{(4)} =  
\bar{N}\left( e^{i \xi}V_0,e^{3 i \xi} u_3^{(3)}\right)
+ N\left( e^{2i \xi}u_2^{(2)},e^{2 i \xi} u_2^{(2)}\right),
\end{align}
\begin{align}
\hat{\mathcal{L}}\left( e^{3 i \xi} u_3^{(5)}\right) &+ \left(4\lambda+\lambda^*\right)  \hat{A} e^{3 i \xi} u_3^{(5)} \nonumber \\
&  = \bar{N}\left( e^{i \xi}V_0,e^{2 i \xi} u_2^{(4)}\right) 
+  \bar{N}\left( e^{-i \xi}V_0^*,e^{4 i \xi} u_4^{(4)}\right)
+  \bar{N}\left( u_0^{(2)} ,e^{3 i \xi} u_3^{(3)}\right) \nonumber \\
& - 3 C_3 e^{3 i \xi} u_3^{(3)},
\end{align}
\begin{align}
\hat{\mathcal{L}}\left( e^{2 i \xi} u_2^{(6)}\right) &+ \left(4\lambda+2\lambda^*\right)  \hat{A} e^{2 i \xi} u_2^{(6)}  \nonumber \\
& = \bar{N}\left( e^{-i \xi}V_0^*,e^{3 i \xi} u_3^{(5)}\right) 
+  \bar{N}\left( u_0^{(2)} ,e^{2 i \xi} u_2^{(4)}\right)
+  \bar{N}\left( u_0^{(4)} ,e^{2 i \xi} u_2^{(2)}\right) \nonumber \\
&+  \bar{N}\left( e^{-2i \xi}u_2^{(2)*},e^{4 i \xi} u_4^{(4)}\right) 
- 2 C_5 e^{2 i \xi} u_2^{(2)}
-  \left( 3 C_3 + C_3^*\right) e^{2 i \xi} u_2^{(4)},
\end{align}
\begin{align}
\hat{\mathcal L} u_0^{(6)} &+ 3\left(\lambda + \lambda^* \right)\hat{A} u_0^{(6)} \nonumber \\
& =\bar{N}\left(u_0^{(2)},u_0^{(4)}\right)
+ \bar{N}\left(e^{-2i\xi}u_2^{(4)*} , e^{2i\xi}u_2^{(2)}\right)
+ \bar{N}\left(e^{-2i\xi}u_2^{(2)*} , e^{2i\xi}u_2^{(4)}\right) \nonumber \\
& + \bar{N}\left(e^{-3i\xi}u_3^{(3)*} , e^{3i\xi}u_3^{(3)}\right)
-\left( C_5 +C_5^*\right) u_0^{(2)}
-2\left( C_3 +C_3^*\right) u_0^{(4)},
\end{align}
which yields
\begin{align}
& C_7 = \frac{1}{\Delta}\biggl\langle e^{i\xi} W_0 \biggl\vert 
   \bar{N}\left(e^{i\xi} V_0, u_0^{(6)}\right) 
+ \bar{N}\left(e^{-i\xi} V_0^*,e^{2 i\xi} u_2^{(6)}\right)
+ \bar{N}\left(e^{-2i\xi} u_2^{(2)*},e^{3 i\xi} u_3^{(5)}\right) \nonumber\\
& \qquad\qquad+ \bar{N}\left(e^{-3i\xi} u_3^{(3)*},e^{4 i\xi} u_4^{(4)}\right)
+ \bar{N}\left(e^{-2i\xi} u_2^{(4)*},e^{3 i\xi} u_3^{(3)}\right)
\biggr\rangle.
\end{align}

To $O\left(\Phi |\Phi|^8\right)$, we have
\begin{align}
\hat{\mathcal{L}}\left( e^{5 i \xi} u_5^{(5)}\right) + 5\lambda  \hat{A} e^{5 i \xi} u_5^{(5)} =  
\bar{N}\left( e^{i \xi}V_0,e^{4 i \xi} u_4^{(4)}\right)
+ \bar{N}\left( e^{2i \xi}u_2^{(2)},e^{3 i \xi} u_3^{(3)}\right),
\end{align}
\begin{align}
\hat{\mathcal{L}}\left( e^{4 i \xi} u_4^{(6)}\right) &+ \left(5\lambda+\lambda^*\right)  \hat{A} e^{4 i \xi} u_4^{(6)} \nonumber \\
& = \bar{N}\left( e^{i \xi}V_0,e^{3 i \xi} u_3^{(5)}\right)
+\bar{N}\left( e^{-i \xi}V_0^*,e^{5 i \xi} u_5^{(5)}\right)
+ \bar{N}\left( u_0^{(2)},e^{4 i \xi} u_4^{(4)}\right) \nonumber \\
&+ \bar{N}\left( e^{2i \xi}u_2^{(2)},e^{2 i \xi} u_2^{(4)}\right)
- 4 C_3 e^{4 i \xi} u_4^{(4)},
\end{align}
\begin{align}
\hat{\mathcal{L}}\left( e^{3 i \xi} u_3^{(7)}\right) &+ \left(5\lambda+2\lambda^*\right)  \hat{A} e^{3 i \xi} u_3^{(7)} \nonumber \\
&  = \bar{N}\left( e^{i \xi}V_0,e^{2 i \xi} u_2^{(6)}\right) 
+  \bar{N}\left( e^{-i \xi}V_0^*,e^{4 i \xi} u_4^{(6)}\right)
+  \bar{N}\left( u_0^{(2)} ,e^{3 i \xi} u_3^{(5)}\right) \nonumber \\
& +  \bar{N}\left( u_0^{(4)} ,e^{3 i \xi} u_3^{(3)}\right) 
+  \bar{N}\left( e^{-2 i \xi}u_2^{(2)*} ,e^{5 i \xi} u_5^{(5)}\right) 
- 3 C_5 e^{3 i \xi} u_3^{(3)} \nonumber \\
& - \left(4 C_3+C_3^*\right) e^{3 i \xi} u_3^{(5)},
\end{align}
\begin{align}
\hat{\mathcal{L}}\left( e^{2 i \xi} u_2^{(8)}\right) &+ \left(5\lambda+3\lambda^*\right)  \hat{A} e^{2 i \xi} u_2^{(8)}  \nonumber \\
& = \bar{N}\left( e^{-i \xi}V_0^*,e^{3 i \xi} u_3^{(7)}\right) 
+  \bar{N}\left( u_0^{(2)} ,e^{2 i \xi} u_2^{(6)}\right)
+  \bar{N}\left( u_0^{(6)} ,e^{2 i \xi} u_2^{(2)}\right) \nonumber \\
&+  \bar{N}\left( e^{-2i \xi}u_2^{(2)*},e^{4 i \xi} u_4^{(6)}\right) 
+  \bar{N}\left( e^{-3i \xi}u_3^{(3)*},e^{5 i \xi} u_5^{(5)}\right) 
+  \bar{N}\left( u_0^{(4)} ,e^{2 i \xi} u_2^{(4)}\right) \nonumber \\
&+  \bar{N}\left( e^{-2 i \xi}u_2^{(4)*},e^{4 i \xi} u_4^{(4)}\right) 
- 2 C_7 e^{2 i \xi} u_2^{(2)}
-  \left( 3 C_5 + C_5^*\right) e^{2 i \xi} u_2^{(4)} \nonumber \\
&-  2\left( 2 C_3 + C_3^*\right) e^{2 i \xi} u_2^{(6)},
\end{align}
\begin{align}
\hat{\mathcal L} u_0^{(8)} &+ 4\left(\lambda + \lambda^* \right)\hat{A} u_0^{(8)} \nonumber \\
& =\bar{N}\left(u_0^{(2)},u_0^{(6)}\right)
+ \bar{N}\left(e^{-2i\xi}u_2^{(6)*} , e^{2i\xi}u_2^{(2)}\right)
+ \bar{N}\left(e^{-2i\xi}u_2^{(2)*} , e^{2i\xi}u_2^{(6)}\right) \nonumber \\
& + \bar{N}\left(e^{-3i\xi}u_3^{(5)*} , e^{3i\xi}u_3^{(3)}\right)
+ \bar{N}\left(e^{-3i\xi}u_3^{(3)*} , e^{3i\xi}u_3^{(5)}\right)
+ \bar{N}\left(e^{-2i\xi}u_2^{(4)*} , e^{2i\xi}u_2^{(4)}\right) \nonumber \\
& + \bar{N}\left(e^{-4i\xi}u_4^{(4)*} , e^{4i\xi}u_4^{(4)}\right) 
+N\left(u_0^{(4)},u_0^{(4)}\right)
-\left( C_7 +C_7^*\right) u_0^{(2)}  \nonumber \\
& -2\left( C_5 +C_5^*\right) u_0^{(4)}
-3\left( C_3 +C_3^*\right) u_0^{(6)},
\end{align}
which yields
\begin{align}
& C_9 = \frac{1}{\Delta}\biggl\langle e^{i\xi} W_0 \biggl\vert 
   \bar{N}\left(e^{i\xi} V_0, u_0^{(8)}\right) 
+ \bar{N}\left(e^{-i\xi} V_0^*,e^{2 i\xi} u_2^{(8)}\right)
+ \bar{N}\left(e^{-2i\xi} u_2^{(2)*},e^{3 i\xi} u_3^{(7)}\right) \nonumber\\
& + \bar{N}\left(e^{-3i\xi} u_3^{(3)*},e^{4 i\xi} u_4^{(6)}\right)
+ \bar{N}\left(e^{-2i\xi} u_2^{(6)*},e^{3 i\xi} u_3^{(3)}\right)
+ \bar{N}\left(e^{-2i\xi} u_2^{(4)*},e^{3 i\xi} u_3^{(5)}\right) \nonumber\\
& + \bar{N}\left(e^{-3i\xi} u_3^{(5)*},e^{4 i\xi} u_4^{(4)}\right)
+ \bar{N}\left(e^{-4i\xi} u_4^{(4)*},e^{5 i\xi} u_5^{(5)}\right)
\biggr\rangle.
\end{align}

To $O\left(\Phi |\Phi|^{10}\right)$, we have
\begin{align}
\hat{\mathcal{L}}\left( e^{6 i \xi} u_6^{(6)}\right) &+ 6\lambda  \hat{A} e^{6 i \xi} u_6^{(6)} =  
\bar{N}\left( e^{i \xi}V_0,e^{5 i \xi} u_5^{(5)}\right)
+ \bar{N}\left( e^{2i \xi}u_2^{(2)},e^{4 i \xi} u_4^{(4)}\right) \nonumber \\
& + N\left( e^{3i \xi}u_3^{(3)},e^{3 i \xi} u_3^{(3)}\right),
\end{align}
\begin{align}
\hat{\mathcal{L}}\left( e^{5 i \xi} u_5^{(7)}\right) &+ \left(6\lambda+\lambda^*\right)  \hat{A} e^{5 i \xi} u_5^{(7)} \nonumber \\
& = \bar{N}\left( e^{i \xi}V_0,e^{4 i \xi} u_4^{(6)}\right)
+ \bar{N}\left( e^{-i \xi}V_0^*,e^{6 i \xi} u_6^{(6)}\right)
+ \bar{N}\left( u_0^{(2)},e^{5 i \xi} u_5^{(5)}\right) \nonumber \\
& + \bar{N}\left( e^{2i \xi}u_2^{(2)},e^{3 i \xi} u_3^{(5)}\right)
+ \bar{N}\left( e^{2i \xi}u_2^{(4)},e^{3 i \xi} u_3^{(3)}\right)
- 5 C_3 e^{5 i \xi} u_5^{(5)},
\end{align}
\begin{align}
\hat{\mathcal{L}}\left( e^{4 i \xi} u_4^{(8)}\right) &+ \left(6\lambda+2\lambda^*\right)  \hat{A} e^{4 i \xi} u_4^{(8)} \nonumber \\
& = \bar{N}\left( e^{i \xi}V_0,e^{3 i \xi} u_3^{(7)}\right)
+\bar{N}\left( e^{-i \xi}V_0^*,e^{5 i \xi} u_5^{(7)}\right)
+ \bar{N}\left( u_0^{(2)},e^{4 i \xi} u_4^{(6)}\right) \nonumber \\
&+ \bar{N}\left( e^{2i \xi}u_2^{(2)},e^{2 i \xi} u_2^{(6)}\right)
+ \bar{N}\left( e^{-2i \xi}u_2^{(2)*},e^{6 i \xi} u_6^{(6)}\right)
+ \bar{N}\left( u_0^{(4)},e^{4 i \xi} u_4^{(4)}\right) \nonumber \\
&+ N\left( e^{2i \xi}u_2^{(4)},e^{2 i \xi} u_2^{(4)}\right)
- 4 C_5 e^{4 i \xi} u_4^{(4)}
- \left(5 C_3+C_3^*\right) e^{4 i \xi} u_4^{(6)},
\end{align}
\begin{align}
\hat{\mathcal{L}}\left( e^{3 i \xi} u_3^{(9)}\right) &+ \left(6\lambda+3\lambda^*\right)  \hat{A} e^{3 i \xi} u_3^{(9)} \nonumber \\
&  = \bar{N}\left( e^{i \xi}V_0,e^{2 i \xi} u_2^{(8)}\right) 
+  \bar{N}\left( e^{-i \xi}V_0^*,e^{4 i \xi} u_4^{(8)}\right)
+  \bar{N}\left( u_0^{(2)} ,e^{3 i \xi} u_3^{(7)}\right) \nonumber \\
& +  \bar{N}\left( u_0^{(6)} ,e^{3 i \xi} u_3^{(3)}\right) 
+  \bar{N}\left( e^{-2 i \xi}u_2^{(2)*} ,e^{5 i \xi} u_5^{(7)}\right) 
+  \bar{N}\left( e^{-3 i \xi}u_3^{(3)*} ,e^{6 i \xi} u_6^{(6)}\right)  \nonumber \\
& +  \bar{N}\left( u_0^{(4)} ,e^{3 i \xi} u_3^{(5)}\right) 
+  \bar{N}\left( e^{-2 i \xi}u_2^{(4)*} ,e^{5 i \xi} u_5^{(5)}\right) 
- 3 C_7 e^{3 i \xi} u_3^{(3)}  \nonumber \\
& - \left(4 C_5+C_5^*\right) e^{3 i \xi} u_3^{(5)}
- \left(5 C_3+2 C_3^*\right) e^{3 i \xi} u_3^{(7)},
\end{align}
\begin{align}
 \hat{\mathcal{L}}\left( e^{2 i \xi} u_2^{(10)}\right) &+ \left(6\lambda+4\lambda^*\right)  \hat{A} e^{2 i \xi} u_2^{(10)}  \nonumber \\
& = \bar{N}\left( e^{-i \xi}V_0^*,e^{3 i \xi} u_3^{(9)}\right) 
+  \bar{N}\left( u_0^{(2)} ,e^{2 i \xi} u_2^{(8)}\right)
+  \bar{N}\left( u_0^{(8)} ,e^{2 i \xi} u_2^{(2)}\right) \nonumber \\
&+  \bar{N}\left( e^{-2i \xi}u_2^{(2)*},e^{4 i \xi} u_4^{(8)}\right) 
+  \bar{N}\left( e^{-3i \xi}u_3^{(3)*},e^{5 i \xi} u_5^{(7)}\right) 
+  \bar{N}\left( u_0^{(4)} ,e^{2 i \xi} u_2^{(6)}\right) \nonumber \\
&+  \bar{N}\left( e^{2 i \xi}u_2^{(4)},u_0^{(6)}\right) 
+  \bar{N}\left( e^{-2 i \xi}u_2^{(4)*},e^{4 i \xi} u_4^{(6)}\right) 
+  \bar{N}\left( e^{-2 i \xi}u_2^{(6)*},e^{4 i \xi} u_4^{(4)}\right)  \nonumber \\
& +  \bar{N}\left( e^{-4 i \xi}u_4^{(4)*},e^{6 i \xi} u_6^{(6)}\right) 
+  \bar{N}\left( e^{-3i \xi}u_3^{(5)*},e^{5 i \xi} u_5^{(7)}\right) 
- 2 C_9 e^{2 i \xi} u_2^{(2)}  \nonumber \\
& -  \left( 3 C_7 + C_7^*\right) e^{2 i \xi} u_2^{(4)}
-  2\left( 2 C_5 + C_5^*\right) e^{2 i \xi} u_2^{(6)}
-  \left( 5 C_3 + 3 C_3^*\right) e^{2 i \xi} u_2^{(8)},
\end{align}
\begin{align}
\hat{\mathcal L} u_0^{(10)} &+ 5\left(\lambda + \lambda^* \right)\hat{A} u_0^{(10)} \nonumber \\
& =\bar{N}\left(u_0^{(2)},u_0^{(8)}\right)
+ \bar{N}\left(e^{-2i\xi}u_2^{(8)*} , e^{2i\xi}u_2^{(2)}\right)
+ \bar{N}\left(e^{-2i\xi}u_2^{(2)*} , e^{2i\xi}u_2^{(8)}\right) \nonumber \\
& + \bar{N}\left(e^{-3i\xi}u_3^{(7)*} , e^{3i\xi}u_3^{(3)}\right)
+ \bar{N}\left(e^{-3i\xi}u_3^{(3)*} , e^{3i\xi}u_3^{(7)}\right)
+ \bar{N}\left(e^{-2i\xi}u_2^{(4)*} , e^{2i\xi}u_2^{(6)}\right) \nonumber \\
&+ \bar{N}\left(e^{-2i\xi}u_2^{(6)*} , e^{2i\xi}u_2^{(4)}\right)
+ \bar{N}\left(e^{-4i\xi}u_4^{(4)*} , e^{4i\xi}u_4^{(6)}\right) 
+ \bar{N}\left(e^{-4i\xi}u_4^{(6)*} , e^{4i\xi}u_4^{(4)}\right) \nonumber \\
&+ \bar{N}\left(e^{-3i\xi}u_3^{(5)*} , e^{3i\xi}u_3^{(5)}\right)
+ \bar{N}\left(e^{-5i\xi}u_5^{(5)*} , e^{5i\xi}u_5^{(5)}\right) 
+ N\left(u_0^{(4)},u_0^{(6)}\right) \nonumber \\
&-\left( C_9 +C_9^*\right) u_0^{(2)} 
-2\left( C_7 +C_7^*\right) u_0^{(4)}
-3\left( C_5 +C_5^*\right) u_0^{(6)} 
-4\left( C_3 +C_3^*\right) u_0^{(8)},
\end{align}
which, finally, yields
\begin{align}
& C_{11} = \frac{1}{\Delta}\biggl\langle e^{i\xi} W_0 \biggl\vert 
   \bar{N}\left(e^{i\xi} V_0, u_0^{(10)}\right) 
+ \bar{N}\left(e^{-i\xi} V_0^*,e^{2 i\xi} u_2^{(10)}\right)
+ \bar{N}\left(e^{-2i\xi} u_2^{(2)*},e^{3 i\xi} u_3^{(9)}\right) \nonumber\\
& + \bar{N}\left(e^{-3i\xi} u_3^{(3)*},e^{4 i\xi} u_4^{(8)}\right)
+ \bar{N}\left(e^{-2i\xi} u_2^{(8)*},e^{3 i\xi} u_3^{(3)}\right)
+ \bar{N}\left(e^{-2i\xi} u_2^{(4)*},e^{3 i\xi} u_3^{(7)}\right) \nonumber\\
& + \bar{N}\left(e^{-3i\xi} u_3^{(7)*},e^{4 i\xi} u_4^{(4)}\right)
+ \bar{N}\left(e^{-4i\xi} u_4^{(4)*},e^{5 i\xi} u_5^{(7)}\right)
+ \bar{N}\left(e^{-2i\xi} u_2^{(6)*},e^{3 i\xi} u_3^{(5)}\right) \nonumber\\
&+ \bar{N}\left(e^{-3i\xi} u_3^{(5)*},e^{4 i\xi} u_4^{(6)}\right)
+ \bar{N}\left(e^{-4i\xi} u_4^{(6)*},e^{5 i\xi} u_5^{(5)}\right)
+ \bar{N}\left(e^{-5i\xi} u_5^{(5)*},e^{6 i\xi} u_6^{(6)}\right)
\biggr\rangle.
\end{align}

\end{appendix}

\begin{acknowledgements}
Discussions with Paulo Arratia are kindly acknowledged. The numerical code used in this work and further research outputs generated through the EPSRC grant EP/I004262/1 can be found at http://dx.doi.org/xxxxxx.
\end{acknowledgements}

\bibliographystyle{spphys}       
\bibliography{references}   

\end{document}